\documentclass[a4paper,11pt]{article}
\usepackage[latin1]{inputenc}
\usepackage{indentfirst}

\usepackage{amsfonts}
\usepackage{float}
\usepackage{latexsym,amssymb,amsmath}
\usepackage[dvips]{graphics,graphicx,epsfig,color}
\usepackage[lofdepth,lotdepth]{subfig}
\begin{document}
\title{\bf Earthquakes in cities revisited\\}
\author{Armand Wirgin\thanks{LMA, CNRS, UPR 7051, Aix-Marseille Univ, Centrale Marseille, F-13453 Marseille Cedex 13, France, ({\tt wirgin@lma.cnrs-mrs.fr})} }
\date{\today}
\maketitle
\begin{abstract}
 During the last twenty years, a  number of  publications of theoretical-numerical nature have appeared which  come to the apparently-reassuring conclusion that  seismic motion on the ground in cities is smaller than what this motion would be in the absence of the buildings (but for the same underground and seismic load). Other than the fact that this finding tells nothing about the motion within the buildings, it must be confronted with the  overwhelming empirical evidence (e.g, earthquakes in Sendai (2011),  Kathmandu (2015), Tainan City (2016), etc.)  that  shaking within buildings of a city is often large enough to damage or even destroy these structures.  I show, on several examples, that  theory can be reconciled  with empirical evidence, and  suggest that the crucial subject of seismic response in cities is in need of more thorough research.
\end{abstract}
\newline
\newline
Abbreviated title: Seismic motion in  cities
\newline
\newline
Corresponding author: Armand Wirgin, tel.: 33 4 91 53 42 43\\ e-mail: wirgin@lma.cnrs-mrs.fr
\newpage
\tableofcontents
\newpage
\newpage
\section{Introduction}\label{intro}
Earthquakes in Mexico City (1985), Kobe (1995), Port-au-Prince (2010), Concepcion (2010), Sendai (2011),  Kathmandu (2015), Tainan City (2016), etc., not to speak of hundreds of earlier destructive seismic events in cities located in dangerous sites (e.g., land-filled portions of, rivers, lakes, seas),  have shown that the  seismic energy transmitted to groups of, and individual, buildings,  can be large enough to damage or even destroy many of these structures.

A widespread representation of why this should happen \cite{r13} is that if a (or group of) building(s)  is in contact with a vibrating ground, then this vibration is communicated to the built structures, all the more easily the closer are the natural frequencies of the buildings  to the soil frequency (assuming of course, that these frequencies are contained within the bandwidth of the the seismic pulse).

This view has given rise in the last twenty years to the idea that since the motion of the ground is the cause (like the diaphragm  of a loudspeaker producing motion of the surrounding air (i.e., sound)), then added mass (i.e., that of the building or group of buildings) to the ground should have the effect of {\it reducing} (or "deamplifying") the ground motion and therefore potentially diminishing the motion communicated to the urban structures \cite{h57}, \cite{ca01}, \cite{ks06}, \cite{br04}, \cite{br06}, \cite{r06}, \cite{bg07}, \cite{tb11a,tb11b}, \cite{gs13}, \cite{it14}, \cite{cl15}, \cite{sn15}. Several researchers \cite{ks06},  \cite{sk08}) have dubbed the building-induced reduction of ground motion a "beneficial effect" of the built component of a site with respect to an incoming seismic wave. For instance,  in \cite{bg07} appears the statement: " Within  a  periodic  city,  effects  of  site  city  interaction  are  beneficial:  the  average ground  motion  is  reduced    (these  results  are  also  valid  when  considering  the building roof motion)".

 Other studies are less-optimistic.   For instance, in  \cite{lc06} it is written that "the peak value of the mean field is slightly smaller..but the mean response decays slower; there is an indication of a potential high value for the total cross-section of resonating buildings".  In the computer-intensive 3D study of \cite{febi06} it is found that, although the ground motion is generally reduced within the city, it is amplified outside of the city.  The authors of  \cite{lc09} find only a slightly smaller ground response at early times, and a slightly higher response at later times. In \cite{dm10} it is written that "the presence of the structure has both the effect of a damper (thus reducing the total energy) and of a filter, focusing energy in the band of building eigenfrequencies". The author of the thesis \cite{t10} writes: "Individual buildings, however, may experience larger or smaller excitation due to ground shaking". The publication \cite{lp10}  evokes "both amplification and deamplification areas",   whereas  in \cite{jbz13}  it is written that the building "mostly reduces mean response",   and in  \cite{it15}  it is stated that the ground motion reduction effect occurs at "high frequencies".

  Pessimistic conclusions are reached in: \cite{jb73},   \cite{gtw05}, \cite{g05}, \cite{wg06a,wg06b}, \cite{dg07},  \cite{gw08}, \cite{md08}, \cite{c08} \cite{pa09}, \cite{skg09},  \cite{gi09}, \cite{g13}, \cite{aa13}, \cite{aa15}. These publications give theoretical-numerical evidence of the fact  that earthquake motion can be {\it stronger} on ground level (at certain frequencies), and more often at higher levels within buildings or groups  of buildings, (especially if the disposition is irregular, and in the vicinity of, or beyond, basin borders  \cite{bg07}), than the ground level motion in the absence of buildings.

 All this points to the need for decisive research  to answer the questions  of:  whether the built component of a city exerts a  positive or negative influence  on the strength of seismic motion therein \cite{h57}, \cite{m08}), and more fundamentally of  why buildings in a city,  at   certain sites and for certain seismic source locations and frequencies, are damaged or destroyed during an earthquake.

My approach to this issue is that: (a) it might be useful to give a status to the building or group of buildings (grouped in what is hereafter termed  "blocks") similar to that of the underlying layers \cite{vw70}, \cite{lw71}, \cite{sw98}, \cite{s98}, \cite{tit01}, \cite{w02}, \cite{dg07}, \cite{md08}; \cite{dm10}, \cite{zs11}), (b)  the generic block should be connected to the latter without discontinuity of displacement and stress so as to fully account for soil-structure interaction \cite{s98,r13},  (c) the focus should be shifted from ground motion to building/block motion since damage to buildings (much less to the ground and underground) is the crucial issue.

I thus show that the energy communicated to the built structures is simply one part of the energy carried by the incoming seismic wave, the other parts being those communicated to the layers and to the half space beneath these layers (the three being connected to the incident energy by the principle of conservation of energy), and that in certain frequency intervals, the energy communicated to a block as a whole, and even to a single building thereof, can be large even when the ground motion is smaller than what it would be in the absence of the block.
\section{Earthquakes at a dangerous site in the absence of  a city}
\begin{figure}[h]
\begin{center}
\includegraphics[width=0.65\textwidth]{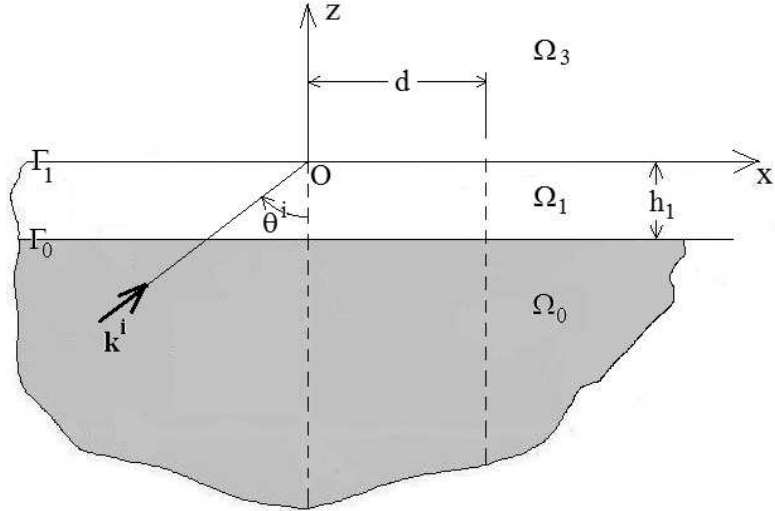}
\caption{Sagittal plane view of a site in the absence of a city. $\Omega_{0}$ is the half plane beneath the layer, $\Omega_{1}$  is the layer, the thickness of which is $h_{1}$, $\Omega_{3}$ is the half space above the layer filled with air  (assumed to be the vacuum), $\Gamma_{0}$ is the interface between the layer and the half-space, $\Gamma_{1}$ is the stress-free ground, $\mathbf{k}^{i}$ and $\theta^{i}$ are the wavevector and incident angle respectively of the seismic plane- body wave disturbance, it being assumed that the motion associated with this wave is in the $y$ direction, $d$ is the width of a representative cell of the site (delimited by the vertical dashed lines).}
\label{sitewithoutcity}
\end{center}
\end{figure}
As in  the "dangerous" sites on which many cities are built, a  half space occupied by a hard, linear, homogeneous, isotropic, lossless medium $M^{[0]}$ is overlain by a series of relatively-soft, linear, homogeneous, isotropic, lossy layers, which I have reduced to a single"equivalent" layer occupied by the relatively-soft, linear, lossy (but dispersion-free) medium $M^{[1]}$, in welded contact with  $M^{[0]}$ \cite{j89}. The  geometrical properties of this site are depicted in fig. \ref{sitewithoutcity}. The incident seismic disturbance is assumed to be a plane body wave whose wavevector $\mathbf{k}^{i}$ lies in the $x-z$ (termed sagittal) plane and the associated motion is perpendicular to the sagittal plane (i.e., the incident wave is SH-polarized).

It is easily shown that the total frequency domain displacement (also, SH-polarized) within, and on the boundaries of, the layer is
\begin{equation}\label{0-010}
u^{[1]}(\mathbf{x},\omega)=2b^{[1]}(\omega)\exp(ik_{x}x)\cos(k_{z}^{[1]}z)~;~\forall \mathbf{x}\in\Omega_{1}~,
\end{equation}
with:
\begin{equation}\label{0-020}
b^{[1]}(\omega)=a^{[0]}(\omega)(e^{[01]}(\omega))^{-1}(-\gamma^{[10]}is^{[11]}(\omega)+c^{[11]}(\omega))^{-1}~,
\end{equation}
$\mathbf{x}$ the vector between the origin $O$ and the point $(x,z)$, $\omega=2\pi f$ the angular frequency, $f$ the frequency, $\theta^{i}$ the incident angle, $s^{i}=\sin\theta^{i}$, $c^{i}=\cos\theta^{i}$, $a^{[0]}(\omega)$ the spectral amplitude of the incident disturbance, $k_{x}=k^{[0]}s^{i}$, $k^{[l]}=\omega/c^{[l]}$, $c^{[l]}= c^{[l]'}+ic^{[l]''}=\sqrt{\mu^{[l]}/\rho^{[l]}}$, $\mu^{[l]}$ the shear modulus and $\rho^{[l]}$ the density in $M^{[l]}$, $k_{z}^{[l]}=\sqrt{(k^{[l]})^2-(k_{x})^{2}}$, $e^{[lm]}(\omega)=\exp(ik^{[l]}h_{m})$, $s^{[lm]}(\omega)=\sin(k^{[l]}h_{m})$, $c^{[lm]}(\omega)=\cos(k^{[l]}h_{m})$, $\gamma^{[lm]}=\mu^{[l]}k_{z}^{[l]}/\mu^{[m]}k_{z}^{[m]}$.

\begin{figure}[ptb]
\begin{center}
\includegraphics[width=0.65\textwidth]{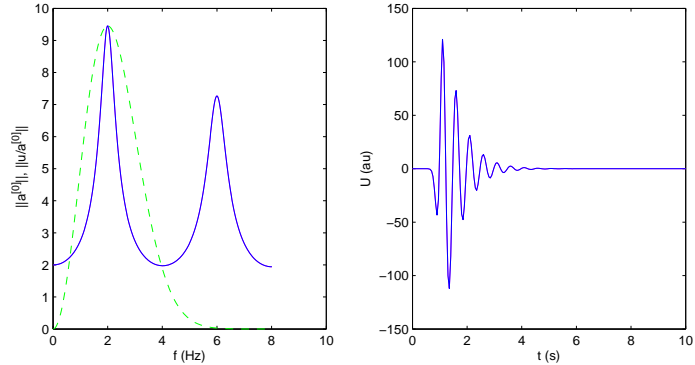}
\caption{Ground motion (i.e, displacement on $z=0$) for the configuration of a single soft layer over a hard half space. Left panel: the blue curve is the modulus of the transfer function, and the dashed green curve denotes the modulus of the Ricker spectrum of the incident pulse. Right panel: the time domain response to the Ricker pulse. $h_{1}=25~m$, $\nu=2~Hz,~\tau=1~s,~\theta^{i}=0^{\circ}$.}
\label{f010}
\end{center}
\end{figure}

It ensues that: (a)  the stress-free surface (i.e., the ground located at $z=0$) is the locus of maximal displacement for all frequencies and (b)  the modulus of the transfer function of ground (i.e., on $z=0$) motion is
\begin{equation}\label{0-030}
\Big\|\frac{u^{[1]}(x,0,\omega)}{a^{[0]}(\omega)}\Big\|=2\Big\|\frac{b^{[1]}(\omega)}{a^{[0]}(\omega)}\Big\|~,
\end{equation}
and if, as is assumed hereafter (i.e., using the parameters of \cite{ks06}; see sect. \ref{kham}), $\big\|\gamma^{[10]}\big\|<1$, then the maxima of the modulus of the ground transfer function occur nearly (i.e., when $\big |c^{[1]''}/c^{[1]'}\big |<<1$) for $s^{[11]}c^{[11]}=0$, or for frequencies
\begin{equation}\label{0-040}
f_{n}=\frac{nc^{[1]'}}{4\eta}~~;~~n=1,3,5,...~.
\end{equation}
wherein $\eta=\frac{h_{1}}{c^{[0]}}\sqrt{\big(\frac{c^{[0]}}{c^{[1]'}}\big)^{2}-\big(s^{i}\big)^{2}}$.
By convention, the so-called soil frequency  of this configuration is obtained for $\theta^{i}=0~\Rightarrow~s^{i}=0$ and $n=1$ and is given by
\begin{equation}\label{0-045}
f_{1}=\frac{c^{[1]'}}{4h_{1}}~,
\end{equation}
which, using the parameters of \cite{ks06}, is equal to $2~Hz$. At this frequency, the modulus of the ground transfer function  takes on the value of $\big\|2/\gamma^{[10]}\big\|$, all this being shown in the left hand panel of fig. \ref{f010} (albeit for a slightly-lossy layer).

Note that the maximum of the transfer function on $z=0$ at the soil frequency is independent of $h_{1}$.
In the left-hand panel of fig. \ref{f010}, I have also depicted $a^{[0]}(\omega)$, i.e., the spectrum of the seismic pulse, assuming, as in \cite{ks06}, that this pulse is Ricker-like, which means that
\begin{equation}\label{0-050}
a^{[0]}(\omega)=\Big(\frac{f}{\nu}\Big)^{2}\exp\Big[i2\pi f\tau-\Big(\Big(\frac{f}{\nu}\Big)^{2}-1\Big)\Big]
~,
\end{equation}
wherein $\nu$ is the frequency at which the pulse is at its maximum (chosen by Kham et al.\cite{ks06} and myself to be $\nu=f_{1}=2 Hz$) and  $\tau$ is related to the onset time of the pulse. Using this information, the time domain displacement on the ground becomes
\begin{equation}\label{0-060}
U^{[l]}(x,0,t)=2\Re\int_{0}^{\infty}u^{[l]}(x,0,\omega)\exp(-i\omega t)d\omega~,
\end{equation}
this function being depicted in the right-hand panel of fig. \ref{f010}.

Not much can be said about these results other than what is well-known to seismologists and earthquake engineers: the presence of a soft layer above, and in welded contact with, a hard half space provokes an amplification \cite{lc06} (assuming, as is the case of a soft layer above a hard layer, that $\|\gamma^{[10]}\|<1$) of the frequency domain seismic response on the ground by a factor $1/\gamma^{[10]}$ at the soil frequency, knowing that this response is equal to 2  at all frequencies when the layer is absent, with the consequence that the time domain response is not only amplified (at its peak value), but also increased in duration by the presence of the soft layer. This is  why I qualified the site as being "dangerous".
\section{Earthquakes at a dangerous site overlain by a uniform city}
%
\subsection{Nature of the  uniform city}\label{unicity}
A city is, by definition, a rather large assembly of buildings. In a modern city, the buildings are grouped into blocks separated by streets for the circulation of people and vehicles.

As is often the case in studies of the effects of earthquakes in cities \cite{ks06}, \cite{bg07}, \cite{gw08}, the buildings are homogenized, which means that their constitutive properties, which vary greatly from one point to another within the buildings, are reduced to average (in some sense) constitutive parameters at all points of the structure. Moreover, the homogenized parameters and geometry vary from one building to another in a given block, so that it proves useful to further homogenize--this time the block-- by assigning an average building height and constitutive parameters to it. A third simplification is to average the block parameters, over the set of blocks and streets of the city, so as to represent the city by a uniform (homogeneous, isotropic, lossy) layer.

\begin{figure}[ptb]
\begin{center}
\includegraphics[width=0.65\textwidth]{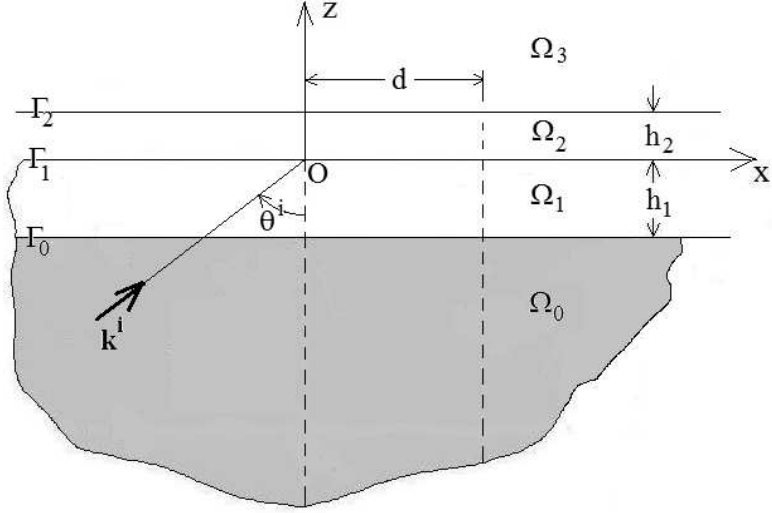}
\caption{Sagittal plane view of a site with the uniform city. $\Omega_{2}$ is the  city layer, filled with the homogeneous, isotropic, lossy medium $M^{[2]}$, The thickness (i.e., height) of the city  is $h_{2}$.  $\Gamma_{2}$ is the stress-free upper surface of the city. Now, $\Gamma_{1}$, which is at ground level, is not stress-free, but rather the locus of continuous stress and displacement. Everything else is as in fig. \ref{sitewithoutcity}.}
\label{sitewithuniformcity}
\end{center}
\end{figure}

The  thus-obtained city (see fig. \ref{sitewithuniformcity}) is here considered as a linear, homogeneous, isotropic, lossy (but dispersion-free)  layer of infinite lateral extent on top of, and in welded contact, with what was formerly the ground (located at $z=0$). This layer is occupied by the medium $M^{[2]}$, its thickness (i.e., later equal to the block height)  is $h_{2}$ and its spatially and spectrally-constant constitutive parameters are $c^{[2]}=c^{[2]'}+ic^{[2]''}$, $\mu^{[2]}$. Now: (a) what was formerly the ground is no longer a stress-free surface, but rather the interface between the city layer and the soil, and (b) it is the top surface of the city that is stress-free.
\subsection{Origin of the so-called beneficial effect of the city}
It is straightforward to show that the frequency domain displacement is:
\begin{equation}\label{0-070}
u^{[2]}(\mathbf{x},\omega)=2a^{[2]}(\omega)\exp(ik_{x}x)\exp(ik_{z}^{[2]}h_{2})\cos\big(k_{z}^{[2]}(z-h_{2})\big)~;~\forall \mathbf{x}\in\Omega_{2}~,
\end{equation}
(and the time domain displacement is given by (\ref{0-060})) with:
\begin{multline}\label{0-080}
a^{[2]}(\omega)=a^{[0]}(\omega)\big(e^{[01]}(\omega)e^{[22]}(\omega)\big)^{-1}\times\\
\big[\gamma^{[21]}is^{[22]}(\omega)\big(-\gamma^{[10]}c^{[11]}(\omega)+is^{[11]}(\omega)\big)+
c^{[22]}(\omega)\big(-\gamma^{[10]}is^{[11]}(\omega)+c^{[11]}(\omega)\big)\big]^{-1}~.
\end{multline}
It ensues that: (a) the "roof" of the city layer (i.e., $z=h_{2}$, which is again the stress-free surface, contrary to the ground $z=0$ which is no longer the stress-free surface) is the locus of maximal displacement for all frequencies, (b) the field at other levels within the city layer can also be maximal, but only at selected frequencies, and c) the modulus of the transfer function of ground (i.e., on $z=0$) motion is
\begin{equation}\label{0-090}
\Big\|\frac{u^{[2]}(x,0,\omega)}{a^{[0]}(\omega)}\Big\|=2\Big\|\frac{a^{[2]}(\omega)c^{[22]}(\omega)}{a^{[0]}(\omega)}\Big\|~,
\end{equation}
Again consider the  situation in which all the media are nearly lossless. Then, at the one-layer frequencies $\omega_{n}=2\pi f_{n}$ defined in (\ref{0-040}),
\begin{equation}\label{0-100}
\Big\|\frac{u^{[2]}(x,0,\omega_{n})}{a^{[0]}(\omega_{n})}\Big\|=\frac{2}{\gamma^{[10]}}
\sqrt{\frac{\big(c^{[22]}(\omega_{n})\big)^{2}}{\big(\gamma^{[21]}\gamma^{[01]}\big)^{2}\big(s^{[22]}(\omega_{n})\big)^{2}+\big(c^{[22]}(\omega_{n})\big)^{2}}}~,
\end{equation}
from which I find
\begin{equation}\label{0-110}
\Big\|\frac{u^{[2]}(x,0,\omega_{n})}{a^{[0]}(\omega_{n})}\Big\|<\Big\|\frac{2}{\gamma^{[10]}}\Big\|~,
\end{equation}
or
\begin{equation}\label{0-120}
\Big\|\frac{u^{[2]}(x,0,\omega_{n})}{a^{[0]}(\omega_{n})}\Big\|<
\Big\|\frac{u^{[1]}(x,0,\omega_{n})}{a^{[0]}(\omega_{n})}\Big\|~,
\end{equation}
wherein
$\Big\|\frac{u^{[1]}(x,0,\omega_{n})}{a^{[0]}(\omega_{n})}\Big\|$ is the transfer function on the ground of the one-layer configuration (i.e., the one in which the city is absent) at the one-layer frequencies $\omega_{n}$. This result is also obtained in the case of lossy layers and means that the presence of a homogeneous layer-like city on top of a given one-layer/half space site {\it reduces the ground motion} at the soil frequency of this site, this being none other than the so-called beneficial effect mentioned in the Introduction.

Note that the maximum of the transfer function on $z=0$ at the soil frequency is now dependent on $h_{1}$ and also on $h_{2}$.
\subsection{Illustration of the reduction of ground motion for a city in the form of a uniform layer}
The city is chosen to have the constitutive  properties of  \cite{ks06}. The city properties are  different from those of the soil, as is likely to occur in reality.  I keep $h_{1}$ constant and increase the thickness $h_{2}$ of the city layer.
\begin{figure}[ht]
\begin{center}
\includegraphics[width=0.65\textwidth]{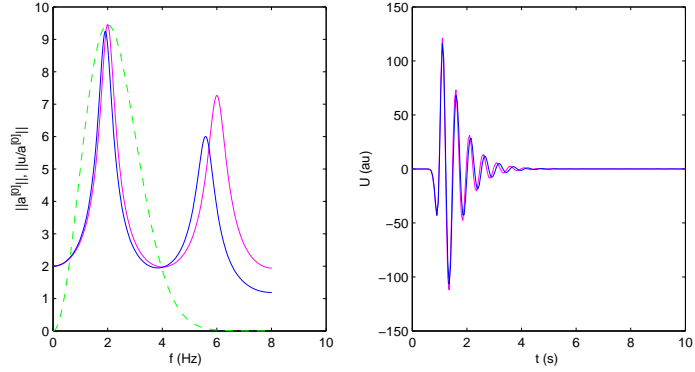}
\caption{Ground motion (i.e., displacement on $z=0$) for the configuration of  two soft layers over a hard half space, the composition of the two layers being different. Left panel: the magenta curve is the modulus of the transfer function in the absence of the uppermost layer, and the blue curve is the modulus of the transfer function in the presence of the uppermost layer, whereas the dashed green curve denotes the modulus of the Ricker spectrum of the incident pulse. Right panel: the time domain responses (same meaning for the colors) to the Ricker pulse. $h_{1}=25~m$, $h_{2}=7.5~m$, $\nu=2~Hz,~\tau=1~s,~\theta^{i}=0^{\circ}$.}
\label{f090}
  \end{center}
\end{figure}
\begin{figure}[ptb]
\begin{center}
      \includegraphics[width=0.65\textwidth]{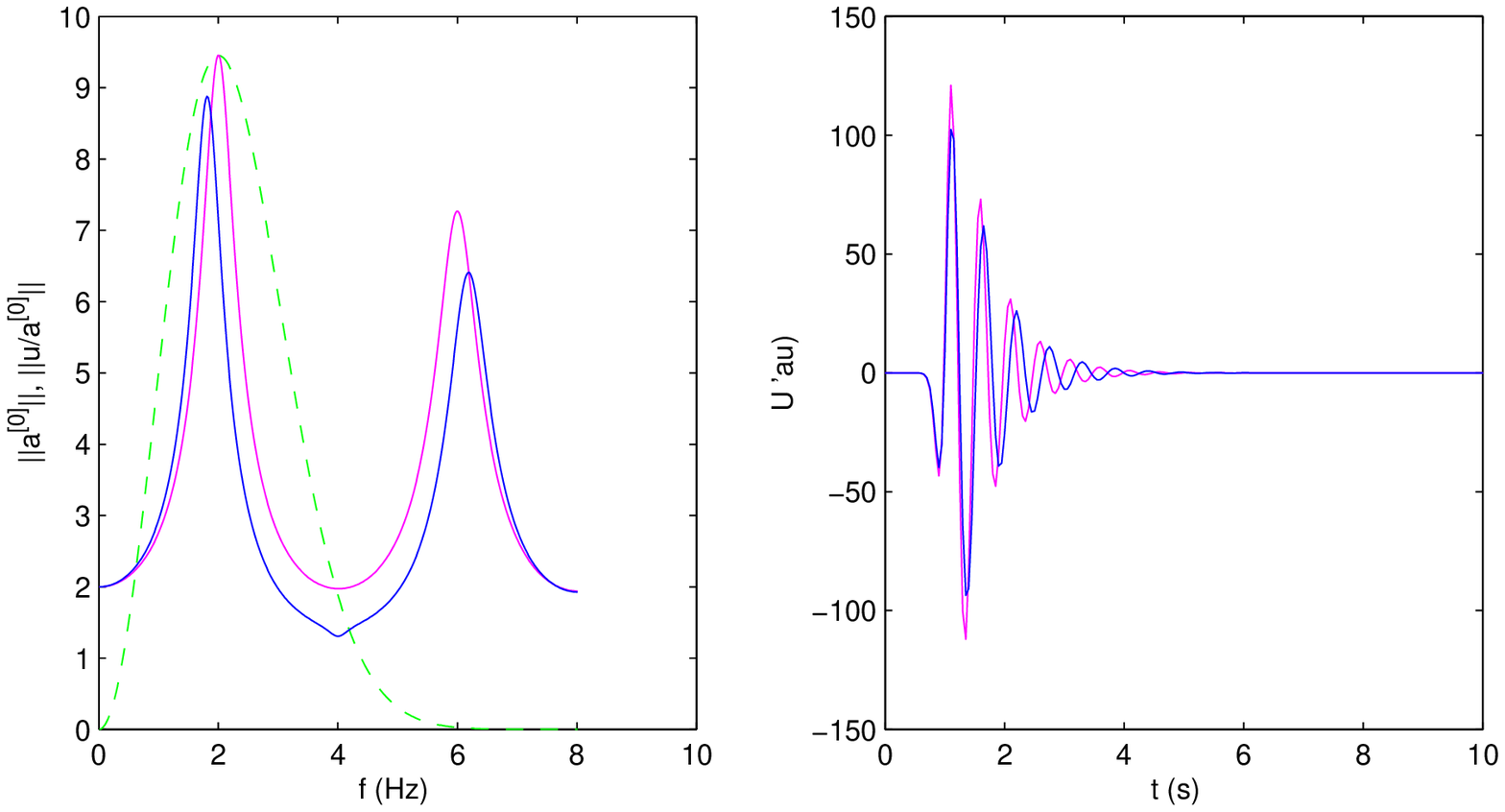}
    \caption{Ground motion (i.e., displacement on $z=0$) for the configuration of  two soft layers over a hard half space, the composition of the two layers being different. Left panel: the magenta curve is the modulus of the transfer function in the absence of the uppermost layer, and the blue curve is the modulus of the transfer function in the presence of the uppermost layer, whereas the dashed green curve denotes the modulus of the Ricker spectrum of the incident pulse. Right panel: the time domain responses (same meaning for the colors) to the Ricker pulse. $h_{1}=25~m$, $h_{2}=15~m$, $\nu=2~Hz,~\tau=1~s,~\theta^{i}=0^{\circ}$.}
    \label{f100}
  \end{center}
\end{figure}
\begin{figure}[ptb]
\begin{center}
      \includegraphics[width=0.55\textwidth]{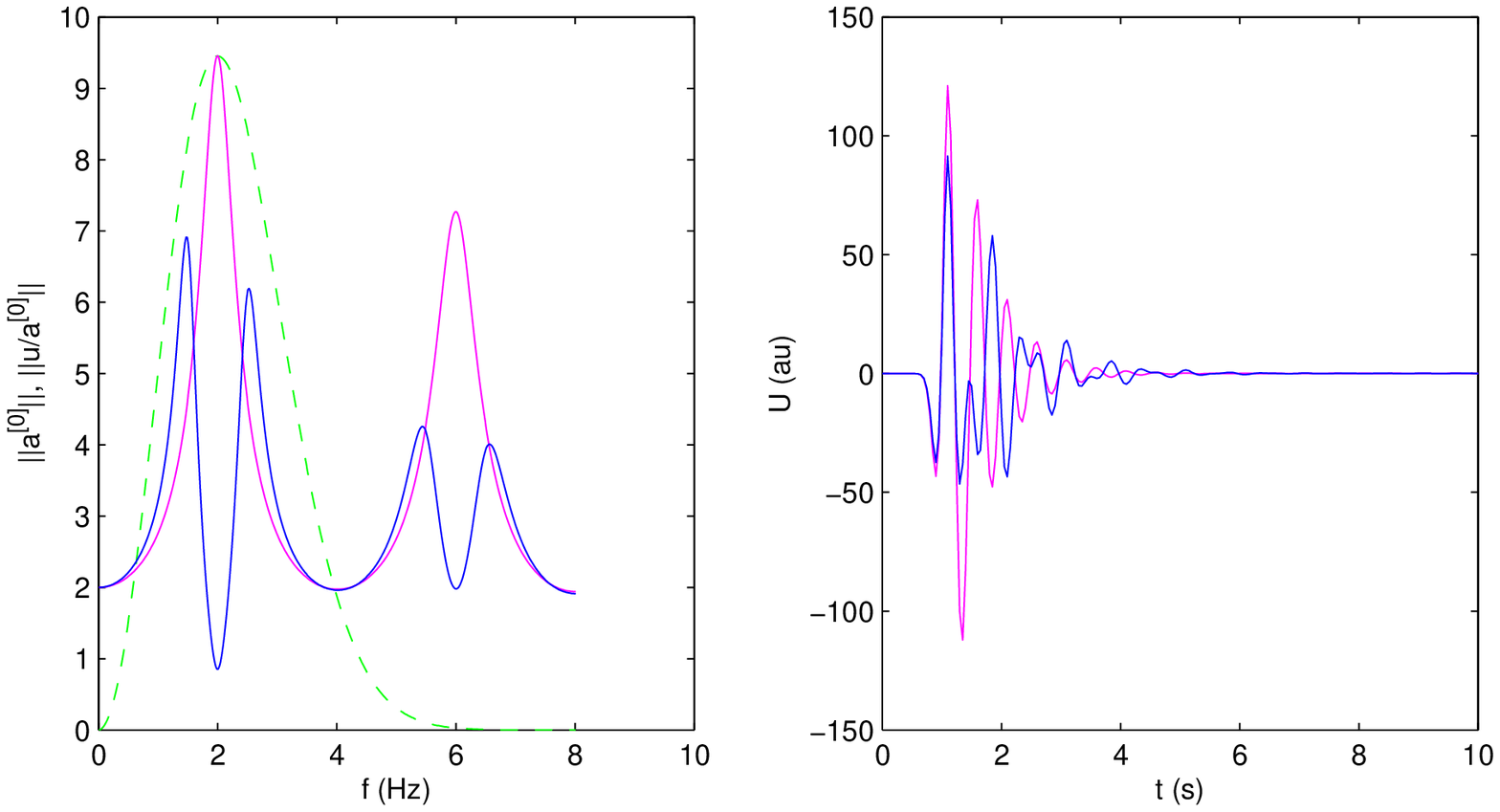}
    \caption{Ground motion (i.e., displacement on $z=0$) for the configuration of  two soft layers over a hard half space, the composition of the two layers being different. Left panel: the magenta curve is the modulus of the transfer function in the absence of the uppermost layer, and the blue curve is the modulus of the transfer function in the presence of the uppermost layer, whereas the dashed green curve denotes the modulus of the Ricker spectrum of the incident pulse. Right panel: the time domain responses (same meaning for the colors) to the Ricker pulse. $h_{1}=25~m$, $h_{2}=30~m$, $\nu=2~Hz,~\tau=1~s,~\theta^{i}=0^{\circ}$.}
    \label{f110}
  \end{center}
\end{figure}
\begin{figure}[ptb]
\begin{center}
      \includegraphics[width=0.55\textwidth]{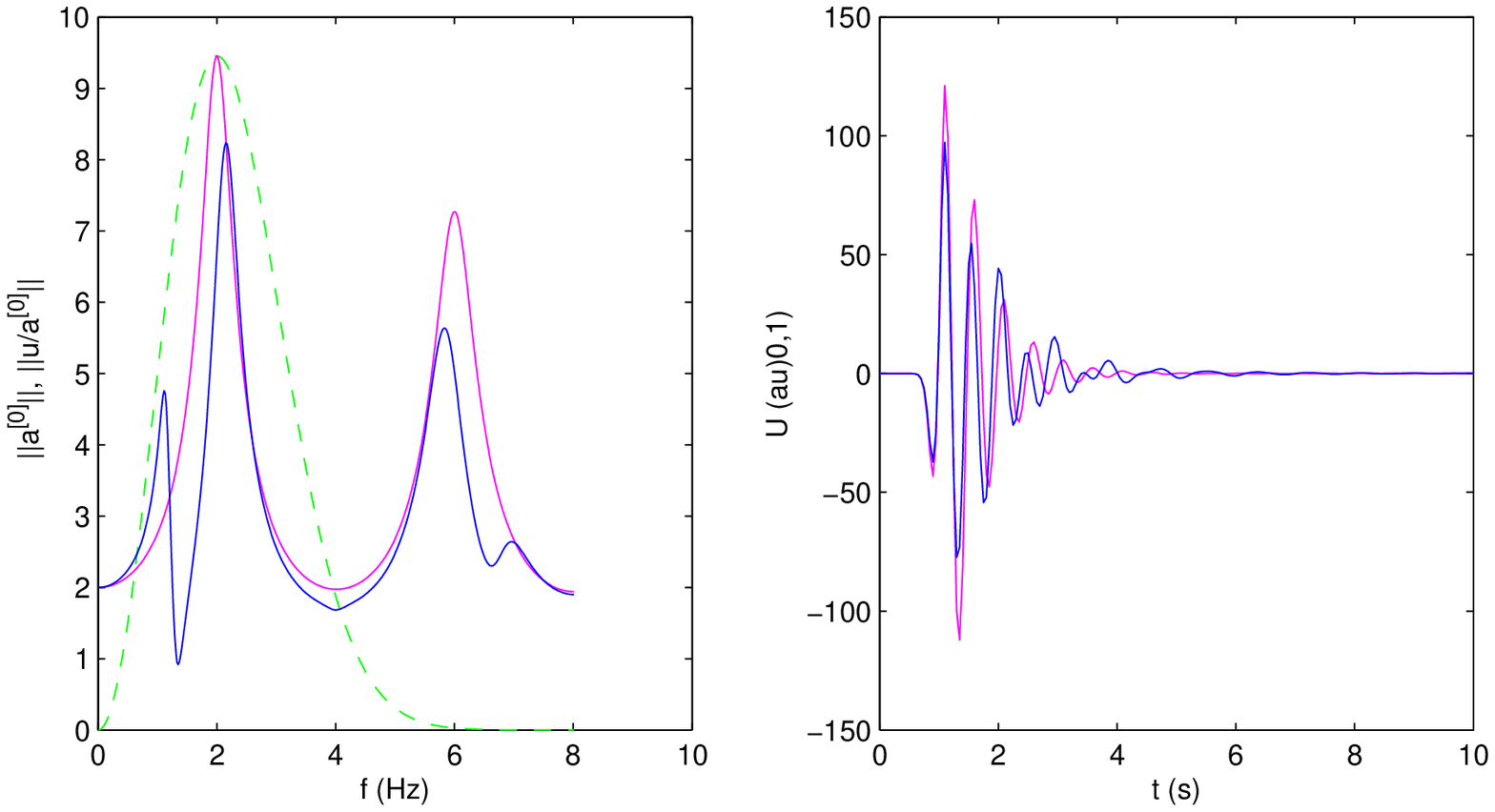}
    \caption{Ground motion (i.e., displacement on $z=0$) for the configuration of  two soft layers over a hard half space, the composition of the two layers being different. Left panel: the magenta curve is the modulus of the transfer function in the absence of the uppermost layer, and the blue curve is the modulus of the transfer function in the presence of the uppermost layer, whereas the dashed green curve denotes the modulus of the Ricker spectrum of the incident pulse. Right panel: the time domain responses (same meaning for the colors) to the Ricker pulse. $h_{1}=25~m$, $h_{2}=45~m$, $\nu=2~Hz,~\tau=1~s,~\theta^{i}=0^{\circ}$.}
    \label{f120}
  \end{center}
\end{figure}
\clearpage
\newpage
\begin{figure}[ptb]
\begin{center}
      \includegraphics[width=0.55\textwidth]{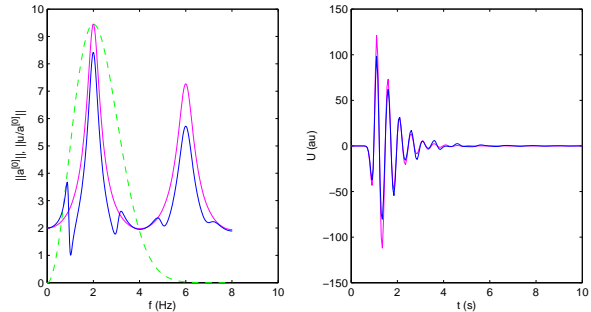}
    \caption{Ground motion (i.e., displacement on $z=0$) for the configuration of  two soft layers over a hard half space, the composition of the two layers being different. Left panel: the magenta curve is the modulus of the transfer function in the absence of the uppermost layer, and the blue curve is the modulus of the transfer function in the presence of the uppermost layer, whereas the dashed green curve denotes the modulus of the Ricker spectrum of the incident pulse. Right panel: the time domain responses (same meaning for the colors) to the Ricker pulse. $h_{1}=25~m$, $h_{2}=60~m$, $\nu=2~Hz,~\tau=1~s,~\theta^{i}=0^{\circ}$.}
    \label{f130}
  \end{center}
\end{figure}

Figs. \ref{f090}-\ref{f130}  tell us what happens at ground level (i.e., $z=0$, which is not the locus of vanishing stress) when I put a homogeneous city layer on the ground and increase the thickness ($h_{1}+h_{2}$, $h_{1}$ constant) of this layer. We again  observe  that the first peak of response on ground level shifts to lower frequency and diminishes in amplitude as the thickness of the overlying layer increases. Again, this produces a reduction of time domain response for constant excitation spectrum (i.e., the Ricker pulse maximum is at $2~Hz$ in these figures). This constitutes  a first example of the  "beneficial effect" of cities when the latter are subjected to an earthquake. This  may  not be convincing since I assumed here that the city was simply a homogeneous layer (albeit of different  composition than of the underlying layer) of infinite extent overlying the ground.

A sidelight of figs. \ref{f090}-\ref{f130} is that for a thickness $h_{2}$ of the city layer greater than $15~m$, the first pseudo-resonance feature is found  to split, as seen previously in \cite{g05,gw08,sb16}  (the latter publications are for cities composed of separated buildings or blocks whereas here I  obtain this effect for a uniform layer city).
\section{Earthquakes in a city composed of identical, uniform, periodically-arranged buildings or blocks: theoretical considerations}
%
\subsection{The grid pattern of many cities}
The pattern of streets in many cities is often  of periodic nature, at least in some areas of the city; this is particularly true in  earthquake-prone cities such as San Francisco, Delhi, and Istanbul. I now study a 2D version (periodicity along the $x$ direction, invariance along the $y$ direction) of such a city.
\subsection{The configuration}
\begin{figure}[ptb]
\begin{center}
\includegraphics[width=12cm] {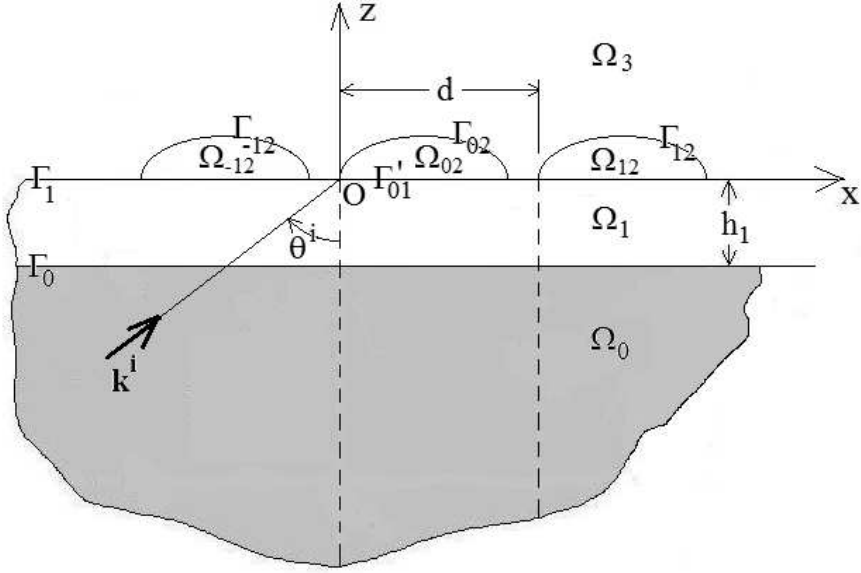}
 \caption{Scattering configuration of a periodic city in the sagittal plane. Normal incidence corresponds to $\theta^{i}=0^{\circ}$.}
  \label{config}
  \end{center}
\end{figure}

The sagittal plane view of the configuration is given in fig. \ref{config}.
Everything is the same as in the uniform city problem except that now the city layer is interrupted by streets so that city layer becomes a periodic (period $d$) set of identical blocks (a block can be composed of a single building, but both the block and building are supposed to be homogeneous) separated by streets. Recall that the periodic nature of the city is suggested by empirical evidence in areas of many modern cities; it is also an efficient means of reducing the complexity (and associated volume of the computations) of the analysis \cite{ks06}, \cite{g05},  \cite{gw08}, \cite{vz14}, \cite{sb15,sb16}.
\subsection{The boundary-value problem}
The following relations (in which the superscripts $+$ and $-$ refer to the upgoing and downgoing  waves respectively) are satisfied by the displacement $u^{[l]}(\mathbf{x},\omega)$ in $\Omega_{l}$:
\begin{equation}\label{1-010}
u^{[l]}(\mathbf{x},\omega)=u^{[l]+}(\mathbf{x},\omega)+u^{[l]-}(\mathbf{x},\omega)~;~l=0,1,2~,
\end{equation}
\begin{equation}\label{1-020}
u_{,xx}^{[l]}(\mathbf{x},\omega)+u_{,zz}^{[l]}(\mathbf{x},\omega)+(k^{[l]})^{2}u^{[l]}(\mathbf{x},\omega)=0~;~\mathbf{x}\in \Omega_{l}~;~l=0,1,2~,
\end{equation}
\begin{equation}\label{1-030}
\mu^{[1]}u_{,z}^{[1]}(\mathbf{x},\omega)=0~;~\mathbf{x}\in \Gamma''_{01}~,
\end{equation}
\begin{equation}\label{1-035}
\mu^{[2]}\boldsymbol{\nu}\cdot\nabla u^{[2]}(\mathbf{x},\omega)=0~;~\mathbf{x}\in \Gamma_{02}~,
\end{equation}
\begin{equation}\label{1-040}
u^{[0]}(\mathbf{x},\omega)-u^{[1]}(\mathbf{x},\omega)=0~;~\mathbf{x}\in \Gamma_{0}~,
\end{equation}
\begin{equation}\label{1-045}
\mu^{[0]}u_{,z}^{[0]}(\mathbf{x},\omega)-\mu^{[1]}u_{,z}^{[1]}(\mathbf{x},\omega)=0~;~\mathbf{x}\in \Gamma_{0}~,
\end{equation}
\begin{equation}\label{1-050}
u^{[1]}(\mathbf{x},\omega)-u^{[2]}(\mathbf{x},\omega)=0~;~\mathbf{x}\in \Gamma'_{01}~,
\end{equation}
\begin{equation}\label{1-055}
\mu^{[1]}u_{,z}^{[0]}(\mathbf{x},\omega)-\mu^{[2]}u_{,z}^{[1]}(\mathbf{x},\omega)=0~;~\mathbf{x}\in \Gamma'_{01}~,
\end{equation}
\begin{equation}\label{1-060}
u^{[0]-}(\mathbf{x},\omega)\sim \text{outgoing waves}~;~\mathbf{x}\rightarrow\infty~,
\end{equation}
wherein $\boldsymbol{\nu}$ is the unit normal vector to $\Gamma_{n2}$, $\Omega_{2}=\bigcup_{n\in\mathbb{Z}}\Omega_{n2}$, $\Gamma_{1}=\bigcup_{n\in\mathbb{Z}}(\Gamma'_{n1}+\Gamma''_{n1})$, $\Gamma'_{n1}$ is the (flat) interface between $\Omega_{1}$ and  $\Omega_{n2}$, $\Gamma''_{n1}$ is the (flat, on $z=0$) portion of the stress-free boundary in the $n$-th cell, $\Gamma_{n2}$ is the  portion above $z=0$ of the stress-free boundary in the $n$-th cell, $u_{,z}$ denotes the partial derivative of $u$ with respect to $z$.

The incident bulk plane wave is
\begin{equation}\label{1-070}
u^{[0]+}(\mathbf{x},\omega)=a^{[0]}(\omega)\exp[i(k_{x}x+k_{z}^{[0]}z)]~.
\end{equation}
The periodicity of the city and the plane wave nature of the solicitation entails (the Floquet condition; $\mathbf{e}_{x}$ is the unit vector along the $x$ axis)
\begin{equation}\label{1-090}
u^{[l]\pm}(\mathbf{x}+d\mathbf{e}_{x},\omega)=u^{[l]\pm}(\mathbf{x},\omega)\exp(ik_{x}^{i}d)~.
\end{equation}
The temporal response $U^{[l]}(\mathbf{x},t)$ is obtained from the spectral response $u^{[l]}(\mathbf{x},\omega)$ via (\ref{0-060}), and
I shall again assume that the spectrum of the seismic solicitation is that of a Ricker pulse.
\subsection{Field representations}
The wave equation (\ref{1-020}), the radiation condition (\ref{1-060}), the Floquet condition (\ref{1-090}), together with the definition (\ref{1-010}) entail:
\begin{equation}\label{1-101}
u^{[l]\pm}(\mathbf{x},\omega)=\sum_{n\in\mathbb{Z}}a_{n}^{[l]\pm}(\omega)\exp[i(k_{xn}x\pm k_{zn}^{[l]}z)]~~;~~l=0,1~,
\end{equation}
wherein
\begin{equation}\label{1-105}
k_{xn}=k_{x}+\frac{2n\pi}{d}~~,~~k_{zn}^{[l]}=\sqrt{(k^{[l]})^{2}-(k_{xn})^{2}}~~;~~ \Re k_{zn}^{[l]}\ge0~,~\Im k_{zn}^{[l]}\ge0~,
\end{equation}
it being understood that $a_{n}^{[0]+}(\omega)=a^{[0]}(\omega)\delta_{n0}$, wherein $\delta_{nm}$ is the Kronecker delta symbol.

For the case of rectangular blocks of width $w$ and height $h_{2}$, an appropriate field representation in the $n=0$ block, satisfying the stress-free boundary condition (\ref{1-035}) on its side walls (assumed to be located at $x=w/2\pm w/2$) and roof (located at $z=h_{2})$, is
\begin{equation}\label{1-111}
u^{[2]\pm}(\mathbf{x},\omega)=\sum_{m=0}^{\infty}a_{m}^{[2]\pm}\cos[K_{xm}x]\exp\big(\pm iK_{zm}^{[2]}(z-h_{2})\big)~,
\end{equation}
with
\begin{equation}\label{1-120}
K_{xm}=\frac{m\pi}{w}~~,~~K_{zm}^{[2]}=\sqrt{(k^{[2]})^{2}-(K_{xm})^{2}}~~;~~ \Re K_{zm}^{[2]}\ge0~,~\Im K_{zm}^{[2]}\ge0~,
\end{equation}
%
\subsection{The exact formulation for rectangular blocks}
For a periodic distribution of rectangular blocks, the unknowns are the five coefficient {\it vectors} $\mathbf{a}^{[0]-}=\{a_{n}^{[0]-}~;~n\in\mathbb{Z}\}$, $\mathbf{a}^{[1]\pm}=\{a_{n}^{[1]\pm}~;~n\in\mathbb{Z}\}$, $\mathbf{a}^{[2]\pm}=\{a_{m}^{[2]\pm}~;~m=0,1,2,...\}$, which must be determined by application of the stress-free condition on the streets (\ref{1-030}) and roofs (\ref{1-035}),  in addition to the continuity of stress and displacement relations at the various interfaces (\ref{1-040})--(\ref{1-055}), as explained in detail in \cite{g05,gw08}. The fields in the various regions are then obtained by inserting these vectors into (\ref{1-101}) and (\ref{1-111}). This is the basis of the rigorous spectral field matching (SFM) formulation.
\subsection{The  approximate formulation for  blocks of arbitrary shape}
The SFM (which is rigorous) is useful for checking approximate formulations, but is not easily generalizable to other polarizations and to 3D problems. I now evoke one such approximate formulations, which has gained a certain favor in studies of soil-building interaction problems.

Let us suppose that the stress-free boundary of the configuration can be represented by the function $z=f(x)=f(x+d)$. For rectangular blocks, $f(x)=h_{2}~;~\forall x\in ]0,w[$ and  $f(x)=0~;~\forall x\in ]w,d[$ in the $n=0$ cell (delimited by dashed vertical lines in fig. \ref{config}). The so-called intersecting canonical body approximation (ICBA)  formulation  \cite{vw70}, \cite{lw71}, \cite{sw98}, \cite{s98}, \cite{w02}) consists first in adopting the following field representations:
\begin{equation}\label{1-130}
u^{[l]\pm}(\mathbf{x},\omega)\approx \tilde{u}^{[l]\pm}(\mathbf{x},\omega)=\tilde{a}_{0}^{[l]\pm}(x,\omega)\exp[i(k_{x0}x\pm k_{z0}^{[l]}z)]~~;~~l=0,1,2~,
\end{equation}
(such that $\tilde{a}_{0}^{[0]+}(x,\omega)=a^{[0]}(\omega)$ and  $\tilde{a}_{0}^{[l]\pm}(x+d,\omega)=\tilde{a}_{0}^{[l]\pm}(x,\omega)\exp(ik_{x}^{i}d)$) and then solving for the five {\it scalar} coefficients $\tilde{a}^{[l-]}(x,\omega)~;~l=0,1,2$ and $\tilde{a}^{[l+]}(x,\omega)~;~l=1,2$ {\it for every $x$} in $[0,d]$ by applying the stress-free boundary condition on $z=f(x)$ and the continuity conditions as before on the flat interfaces located at $z=-h_{1}$ and $z=0$.
The final step in this procedure is to insert these coefficients into (\ref{1-130}) so as to obtain the fields in the three regions of the $n=0$ cell of the configuration.

The ICBA is no longer an approximate formulation (i.e., is exact) when the city takes the form of a homogeneous layer of constant thickness and infinite extent.
\subsection{Exact expression of the conservation of energy}\label{exen}
The demonstration of what becomes of seismic {\it ground motion}  when buildings are placed on the ground does not answer the question of what becomes of {\it building motion} under seismic loading. We know that some of these buildings will be damaged or destroyed which means that they are necessarily the recipients of a part of the incident energy. How, and how much of, this energy is injected into the buildings is the as yet unanswered question.

The first step towards an answer is to obtain the fields in the half space, layer and buildings by applying the boundary and interface conditions, in either the SFM or ICBA formulations, as explained in the previous sections. The second step, which I now address, is to establish an expression of the conservation of energy, relating the incident energy to the energies distributed to the component areas of the scattering configuration.

Eqs. (\ref{1-010})-(\ref{1-020}) lead to
\begin{equation}\label{ce0}
\int_{\Omega_{l}}[u^{[l]*}\Delta u^{[l]}-u^{[l]}\Delta u^{[l]*}]d\varpi+
\Im\big[(k^{[l]})^{2}\big]\int_{\Omega_{l}}\|u^{[l]}\|^{2}d\varpi=0~,
\end{equation}

Applying Green's second identity in the $n=0$ cell of the configuration, and remembering (\ref{1-070}) and (\ref{1-101}),  leads rigorously to the conservation of energy relation
\begin{equation}\label{ce1}
\Re\Big[\sum_{n\in\mathbb{Z}}\|a_{n}^{[0]-}\|^{2}k_{zn}^{[0]}d\Big]+\sum_{l=1}^{2}\frac{\mu^{[l]}}{\mu^{[0]}}\Im\big[(k^{[l])^{2}}\big]
\int_{\Omega_{0l}}\|u^{[l]}\|^{2}d\varpi=\|a^{[0]}\|^{2}k_{z}^{[0]}d~,
\end{equation}
($d\varpi$ is the differential area element in the sagittal plane and $\Omega_{0l}$ is the  $n=0$ cell portion of $\Omega_{l}$) or, after normalization:
\begin{equation}\label{ce2}
\rho(\omega)+\sum_{l=1}^{2}\alpha_{l}(\omega)=1~,
\end{equation}
wherein 
\begin{equation}\label{ce3}
\rho(\omega)=\Re\Big[\sum_{n\in\mathbb{Z}}\frac{\|a_{n}^{[0]-}\|^{2}}{\|a^{[0]}\|^{2}}\frac{k_{zn}^{[0]}}{k_{z}^{[0]}}\Big]~~,~~ \alpha_{l}(\omega)=\frac{\mu^{[l]}}{\mu^{[0]}}\frac{\Im\big[(k^{[l])^{2}}\big]}{\|a^{[0]}\|^{2}k_{z}^{[0]}}\int_{\Omega_{0l}}\|u^{[l]}\|^{2}d\varpi~,
\end{equation}
 with $\rho$  the normalized total reflected energy (into the half space, this being often called the radiation damping term \cite{ws96}), and  $\alpha_{l}$ the normalized absorbed energy in the domain $\Omega_{0l}$ occupied by medium $M^{[l]}$, it being recalled that $M^{[0]}$ is non-dissipative and the $n=0$  block of the city occupies $\Omega_{02}$. This means that the larger is $\alpha_{2}$, the more of the input energy is transferred into the city (and the less is the fraction of input energy transferred to the layer and/or to radiation damping).

 Eq. (\ref{ce2}), whose  right-hand term is the normalized input energy (furnished by the incident seismic wave) and left-hand term the normalized output energy, is the rigorous expression of the conservation of energy
\subsection{Expression of the conservation of energy in the approximate formulation}\label{apen}
Employing the ICBA instead of the rigorous fields requires a reinterpretation of (\ref{ce1})-(\ref{ce3}). To illustrate this, I refer back to the situation of rectangular blocks. In the $n=0$ cell (composed of subregions $\Omega_{0l}~;~l=0,1,2$), the block occupies $\Omega'_{02}=\Omega_{02}$. Beneath the block (i.e., for $x\in [0,w]$) are situated the subregions $\Omega'_{01}\subset\Omega_{01}$ and $\Omega'_{00}\subset\Omega_{00}$. To the right of the block, and beneath $\Gamma''_{01}$ (i.e., for $x\in [w,d]$), are situated the subregions $\Omega''_{01}\subset\Omega_{01}$ and $\Omega''_{00}\subset\Omega_{00}$. The ICBA leads to considering what happens in the column $\Gamma'_{01}|\Omega'_{02}|\Omega'_{01}|\Omega'_{00}$ as a three layer (each of infinite lateral extent) problem; likewise, it leads to considering what happens in the column $\Gamma''_{01}|\Omega''_{01}|\Omega''_{00}$ as a two layer (each of infinite lateral extent) problem. Let the solution of the first of these two problems be $\tilde{u}^{[l]\pm'}$ to which is associated $\tilde{a}^{[l]\pm'}$ and let the solution of the second of these two problems be $\tilde{u}^{[l]\pm''}$ to which is associated $\tilde{a}^{[l]\pm''}$. Then, I find it  appropriate to write the conservation of energy relation as
\begin{equation}\label{ce4}
\rho'(\omega)+\sum_{l=1}^{2}\alpha'_{l}(\omega)+\rho''(\omega)+\sum_{l=1}^{2}\alpha''_{l}(\omega)=1~,
\end{equation}
wherein
\begin{equation}\label{ce5}
\rho'(\omega)=\frac{\|\tilde{a}_{0}^{[0]-'}\|^{2}}{\|a^{[0]}\|^{2}}~~,~~ \alpha'_{l}(\omega)=\frac{\mu^{[l]}}{\mu^{[0]}}\frac{\Im\big[(k^{[l])^{2}}\big]}{\|a^{[0]}\|^{2}k_{z}^{[0]}}
\int_{\Omega'_{0l}}\|\tilde{u}^{[l]'}\|^{2}d\varpi~,
\end{equation}
\begin{equation}\label{ce6}
\rho''(\omega)=\frac{\|\tilde{a}_{0}^{[0]-''}\|^{2}}{\|a^{[0]}\|^{2}}~~,~~ \alpha''{l}(\omega)=\frac{\mu^{[l]}}{\mu^{[0]}}\frac{\Im\big[(k^{[l])^{2}}\big]}{\|a^{[0]}\|^{2}k_{z}^{[0]}}
\int_{\Omega''_{0l}}\|\tilde{u}^{[l]''}\|^{2}d\varpi~.
\end{equation}
\clearpage
\newpage
\section{Earthquakes ~in ~a ~city ~composed ~of ~identical, homogenized, periodically-arranged buildings: numerical results}
%
\subsection{Analysis of the Kham et al. results for a periodic block city}\label{kham}
I refer here to the article \cite{ks06} and, in particular to the results therein relative to a periodic city composed of identical B2S (rectangular, homogenized) blocks (qualified as buildings, and separated by streets, in \cite{ks06})  overlying a soft layer (actually a trapezoidal basin $500m$ wide, but transformed by me to a layer of infinite lateral extent) underlain by a hard half space. The parameters (in my notation) are:\\\\
\\
$h_{1}=25 ~m$\\
$h_{2}=30 ~m$\\
$w=10 ~m$\\
$c^{[0]'}=1000 ~ms^{-1}$\\
$Q^{[0}=100$\\
$\rho^{[0]}=2000 ~kg~m^{-3}$\\
$c^{[1]'}=200 ~ms^{-1}$\\
$Q^{[1]}=25$\\
$\rho^{[1]}=1800 ~kg~m^{-3}$\\
$f_{b}^{[2]}=2 ~Hz$\\
$Q^{[2]}=10$\\
$\rho^{[2]}=250 ~kg~m^{-3}$\\
$\theta^{i}=0^{\circ}$\\
$\nu=2~Hz$\\
$\tau=1~s$.\\\\
To convert this set of parameters to my own set of parameters, I first make use of the definition \cite{c15} of the quality factor $Q$ relative to a complex velocity $c=c'+ic''$ (with $c'\geq0$ and $c''\leq 0$:
\begin{equation}\label{ks1}
 Q=-\frac{\Re (c^{2})}{\Im( c^{2})}=-\frac{(c')^{2}-(c'')^{2}}{2c'c''}~,
\end{equation}
so that, for $|c''|<<|c'|$
\begin{equation}\label{ks2}
 Q\approx-\frac{(c')^{2}}{2c'c''}=-\frac{c'}{2c''}~,
\end{equation}
whence
\begin{equation}\label{ks3}
 c''\approx-\frac{c'}{2Q}=c'\zeta~,
\end{equation}
wherein $\zeta=\frac{1}{2Q}$ is the damping ratio.

For a fixed-base shear wall (the $m=0$ mode is dominant in (\ref{1-111}))  of height $h$, the displacement at its base $z=0$ is nil, so that the frequency $f$ must obey
\begin{equation}\label{ks4}
\cos\left(\frac{2\pi f}{c'}h\right)=0~,
\end{equation}
and the lowest frequency for this to occur obeys
\begin{equation}\label{ks5}
\frac{2\pi f_{b}}{c'}h=\frac{\pi}{2}~,
\end{equation}
or
\begin{equation}\label{ks5}
f_{b}=\frac{c}{4h}~,
\end{equation}
which is frequently termed the (first) {\it building frequency}.

Lastly, the velocity $c^{[l]}$  is related to the density $\rho^{[l]}$ and shear modulus $\mu^{[l]}$ therein by
\begin{equation}\label{ks7}
\mu^{[l]}=\rho (c^{[l]})^{2}~.
\end{equation}
so that  the  employment of  (\ref{ks3}), (\ref{ks5}), (\ref{ks7}) (neglecting the imaginary parts therein), together with the parameters $Q^{[l]},~f_{b}, ~\rho^{[l]}$ in \cite{ks06}), gives rise to:\\\\
$c^{[0]''}=-1000/200=-5\approx 0~ms^{-1}$\\
$\mu^{[0]}=2000\times 10^{6}=2\times 10^{9} Pa$\\
$c^{[1]''}=-200/50=-4~ms^{-1}$\\
$\mu^{[1]}=1800\times 4\times 10^{4}=7.2\times 10^{7} Pa$\\
$c^{[2]'}=2\times 120=240~ms^{-1}$\\
$c^{[2]''}=-\frac{240}{20}=-12~ms^{-1}$\\
$\mu^{[2]}=250\times (240)^{2}=14.4\times 10^{6} Pa$.\\

In \cite{ks06}, the number $N_{b}$ of buildings (of width $w$) is varied for a fixed basin width $L=500 m$, it being understood that the left-most side of the first building is at $x=0$ and the right-most side of the last building is at $x=L$. The problem is to determine the interval $s$ between successive buildings, and to do this, I make use of
\begin{equation}\label{ks8}
N_{b}w+(N_{b}-1)s=L~.
\end{equation}
Furthermore, since, the parameter that enters into my formulation is $d$ (the distance along $x$) between two successive building midpoints and these two buildings have (in the present instance) the same width $w$, I have $d=s+w$. Thus, for instance, when $N_{b}=25$, $s=\frac{500-10\times 25}{24}=10.42$ so that $d=20.42~m$.

\begin{figure}[ht]
  \begin{center}
      \includegraphics[width=0.8\textwidth]{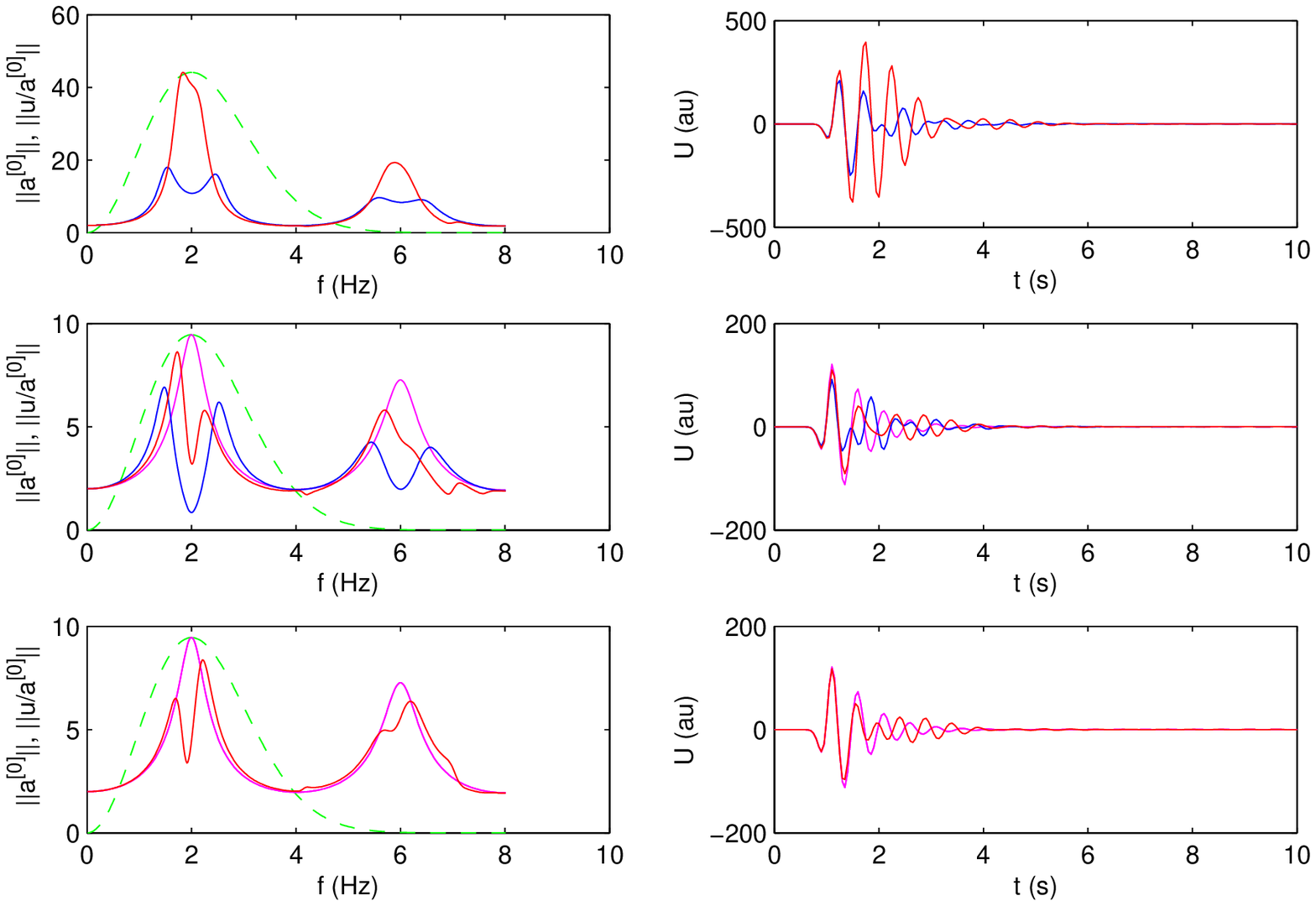}
    \caption{Left-hand panels: modulus of transfer functions of displacement at the roof (top panels), base (middle panels), and on the ground in-betwwen buildings (bottom panels). Right-hand panels: corresponding time domain displacement responses. $N_{b}=10$. $d=54.44~m$. $w_{2}=10~m$. $h_{2}=30~m$. $\nu=2~Hz,~\tau=1~s,~\theta^{i}=0^{\circ}$.}
    \label{fks3}
  \end{center}
\end{figure}

\begin{figure}[ptb]
  \begin{center}
      \includegraphics[width=0.8\textwidth]{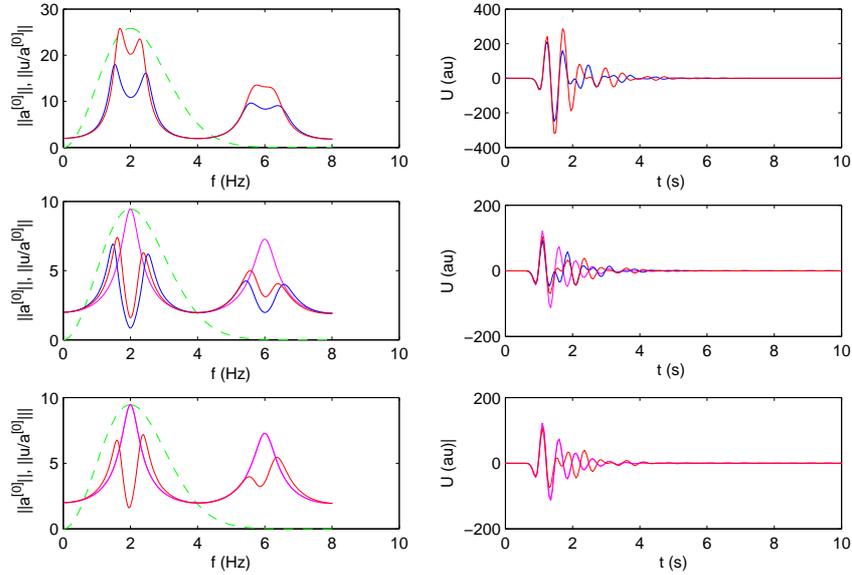}
    \caption{Left-hand panels: modulus of transfer functions of displacement at the roof (top panels), base (middle panels), and on the ground in-between buildings (bottom panels). Right-hand panels: corresponding time domain displacement responses. $N_{b}=25$. $d=20.42~m$. $w_{2}=10~m$. $h_{2}=30~m$.  $\nu=2~Hz,~\tau=1~s,~\theta^{i}=0^{\circ}$.}
    \label{fks5}
  \end{center}
\end{figure}

The results of my computations are given in figs. \ref{fks3}-\ref{fks5}. Each of these figures is composed of three rows and two columns of panels. The first, second, and third rows are relative to motion at the midpoint of the generic building roof, midpoint of the building base and midpoint of the interbuilding ground-level segment respectively. In the left-hand panels, the magenta curve is the modulus of the transfer function in the absence of the city (this is often termed the {\it free-field transfer function}),  the blue curve  the modulus of the transfer function in the presence of the city  blocks resulting from the ICBA formulation, and the red curve the modulus of the transfer function in the presence of the city  blocks resulting from the SFM formulation, whereas the dashed green curve denotes the modulus of the Ricker spectrum of the incident pulse. The right-hand panels refer to the time domain responses to the Ricker pulse, with the same meaning for the colors as in the left-hand panels.
\clearpage
\newpage

The results related to the red curves in figs. \ref{fks3}-\ref{fks5}, obtained here by the rigorous SFM method explained in detail in \cite{g05}, \cite{gw08}), as well as others not shown,  are, in fact, identical to  the corresponding results in \cite{ks06} obtained by the (rigorous) boundary element method,  and call for the following comments:\\\\
(a) they apply to increasing city density ($\propto N_{b}/L$);\\
(b) at the building frequency, the transfer function on the ground in between buildings attains its minimum value whereas the free-field transfer function attains its maximum, this being true for all city densities;\\
(c) The maximum of the transfer function at the building roof is systematically greater than its counterpart at the base of the building;\\
(d) the maximum of the transfer function at the base of the building is systematically smaller (but not much so) than the maximum of the free-field transfer function;\\
(e) At the building frequency, the minimum of the transfer function at the base of the building does not vanish as it would if the building really responded as a fixed-base shear wall;\\
(f) The maximum of the transfer function at the roof of the building is systematically larger (by a factor of the order of 3) than the maximum of the free-field transfer function;\\
(g)  The peak temporal response on the ground between the buildings is systematically smaller (but not much so, except for the least dense city for which it is slightly larger) than the peak temporal free-field  ground level response;\\
(h) The peak temporal response at the base of a building is systematically smaller (but not much so) than the peak temporal free-field ground level response;\\
(i) the durations of the response signals are rather short (of the order of $5~s$) both in, and in between, the buildings;\\
(j) there appears a little bit of beating in the response signals, particularly for the less dense cities.
\\\\\
It can be observed in these figures that the ICBA (blue curves) generally predicts a smaller response than the rigorous formulation, particularly at the roof of the buildings; I shall show later on that this is a specific feature of relatively-large aspect ratio ($h_{2}/w>1$) structures (i.e., buildings) since the ICBA agrees quite well with the rigorous formulation for blocks ($h_{2}/w<1$).

From observations (d) and (g), and other similar results, Kham et al. conclude \cite{ks06} "A so-called site-city resonance is reached when the soil fundamental frequency and structure eigenfrequencies coincide; building vibrations and ground motion are then significantly decreased...".

This is an illustration of the existence of a "beneficial effect" in the case of a city composed of a periodic set of homogeneous buildings  (recall that earlier I showed that similar effects concerning ground motion are produced in uniform cities). Although my rigorous results support the Kham et al.  statement as concerns ground motion, I cannot agree with what they write about the decrease of building vibrations because if the "significant decrease" is with respect to frequency than figs. \ref{fks3} and \ref{fks5}  show that the roof motion is not minimal at the soil fundamental frequency and if the "significant decrease" is with respect to the situation in which there are no buildings then the statement is meaningless since there can be no building vibrations in the absence of buildings.  However, if this reduction of building vibrations is understood by comparison to what occurs in a single building as in \cite{sn15}, then my  objection is unfounded (see also \cite{gw08}).

The question is now: if cities (whether uniform or composed of buildings separated by streets)  indeed produce "beneficial" (in the sense explained above) seismic effects during earthquakes, then why are the built structures in these cities damaged or destroyed during earthquakes? I shall show hereafter that the "beneficial effect" refers only to locations on the ground (on the streets or perhaps at the base of the blocks) and only at the soil frequency, whereas it is the building (i.e., not the ground) as a whole (and even the blocks of which it is a part), at all the frequencies within the bandwidth of the seismic spectrum, that should retain our attention.
\subsection{The energies communicated to the half space, layer and buildings in the Kham et al. examples}\label{khamen}
\begin{figure}[ht]
  \begin{center}
      \includegraphics[width=0.8\textwidth]{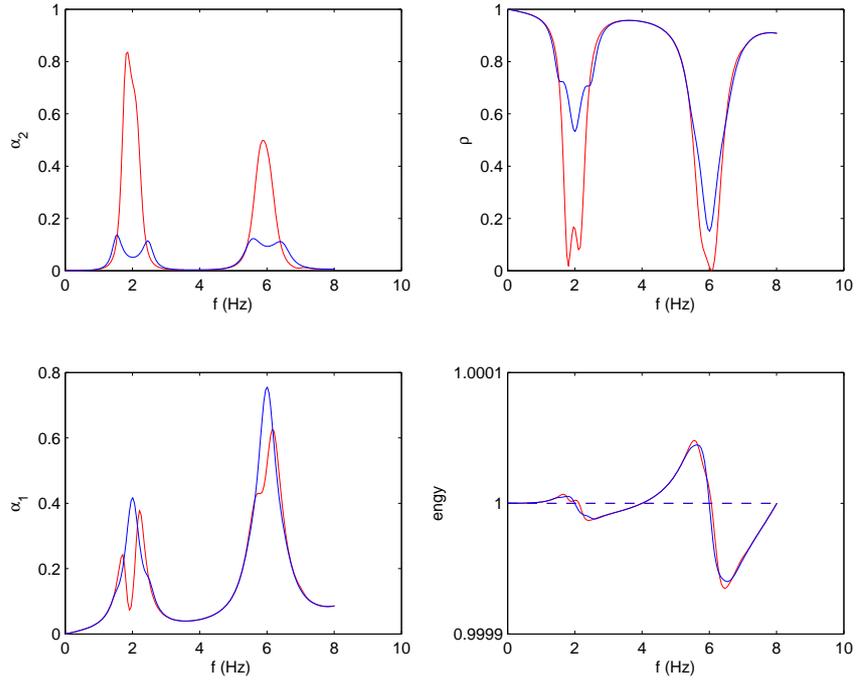}
    \caption{The upper left-hand panel pertains to  the spectrum of normalized energy  communicated to the generic building ($\alpha_{2}(\omega)$). The lower left-hand panel depicts the spectrum of normalized energy communicated to the soft layer ($\alpha_{1}(\omega))$. The upper right-hand panel pertains to the spectrum of normalized energy communicated to the hard half space (i.e., radiation damping). The lower right-hand panel depicts the spectra of normalized input energy  (dashed curves) and normalized output energy (full curves). $N_{b}=10$. $d=54.44~m$. $w_{2}=10~m$. $h_{2}=30~m$. $\theta^{i}=0^{\circ}$.}
    \label{fks7}
  \end{center}
\end{figure}

\begin{figure}[ptb]
  \begin{center}
      \includegraphics[width=0.8\textwidth]{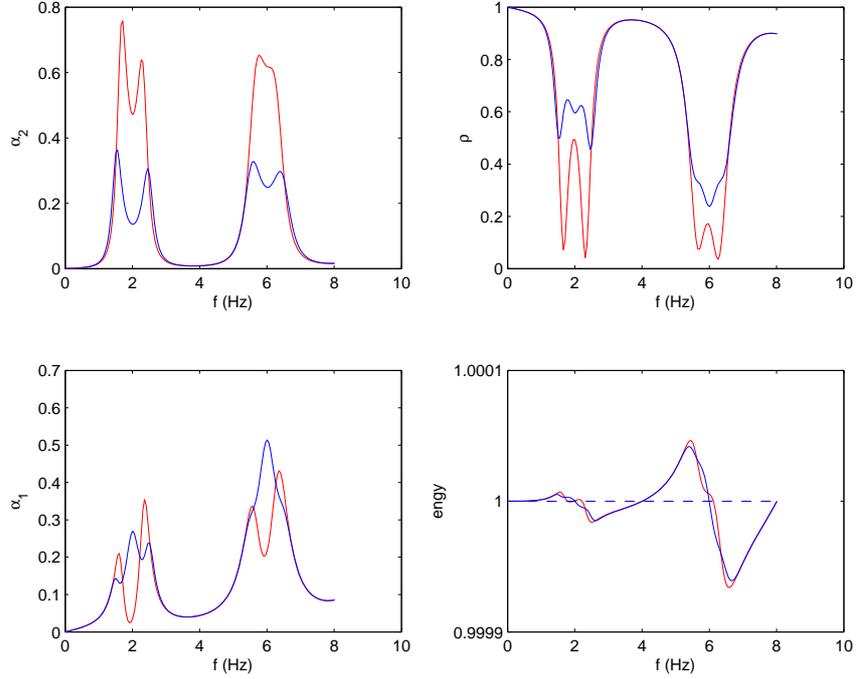}
    \caption{The upper left-hand panel pertains to  the spectrum of normalized energy  communicated to the generic building ($\alpha_{2}(\omega)$). The lower left-hand panel depicts the spectrum of normalized energy communicated to the soft layer ($\alpha_{1}(\omega))$. The upper right-hand panel pertains to the spectrum of normalized energy communicated to the hard half space (i.e., radiation damping). The lower right-hand panel depicts the spectra of normalized input energy  (dashed curves) and normalized output energy (full curves).  $N_{b}=25$. $d=20.42$. $w_{2}=10$. $h_{2}=30$. $\theta^{i}=0^{\circ}$.}
    \label{fks8}
  \end{center}
\end{figure}

To make the last statement of sect. \ref{kham} more tangible, I computed the various terms in the conservation of energy relations of sects. \ref{exen} and \ref{apen}. Figs. \ref{fks7} and  \ref{fks8} are relative to figs. \ref{fks3} and  \ref{fks5} respectively. The red energy spectra curves were obtained via the rigorous SFM fields, whereas the blue energy spectra curves result from the ICBA approximate fields.
\clearpage
\newpage

These figures show clearly that a substantial part of the incident seismic energy is communicated to the buildings near the soil frequency (in this case, $2~Hz)$. A lesser fraction is communicated to the layer, and even less is devolved to radiation damping in the neighborhood of the soil frequency. It thus appears that the buildings of a city, of characteristic frequency close to that of the soil, are likely to be damaged or destroyed when the maximum of the spectrum of the incoming seismic pulse is near  the soil frequency, this being so because at this frequency, the major part of the incident energy is communicated to these buildings.

A sidelight of figs.  \ref{fks7}~ and  \ref{fks8}  is that a substantial amount of energy is transferred to the buildings even at the second building frequency (6~Hz) thus making it also possible for large damage to be inflicted to the buildings when the major part of the incident energy is in a band centered on this frequency.

A second sidelight of figs.  \ref{fks7} and  \ref{fks8} is that the ICBA also supports the previous two statements, but underestimates the proportion of energy communicated to the buildings near the soil frequencies.
\clearpage
\newpage
\section{Earthquakes ~in~ a ~city ~composed ~of identical, homogeneous, periodically-arranged blocks: numerical results}
%
\subsection{Preliminaries}
The configurations studied by Kham et al. \cite{ks06} and many others consisting either of periodic assemblies  of isolated buildings \cite{wb96,ca01,g05,gw08,wg06a,wg06b,br06,sn15,sb15,sb16} or  more-or-less randomly-arranged sets \cite{ca01}, \cite{lc06}, \cite{lc09}, \cite{febi06}, \cite{gtw05}, \cite{it14}, \cite{it15}, \cite{lc09}, \cite{pa09}, \cite{t10}, \cite{tb11a}, \cite{tb11b}, (tens of elements) of isolated buildings are perhaps not the most representative examples of what a real city  looks like because the latter is usually not periodic at the scale of a building, nor is it random (in certain areas of more-modern cities) at the scale of a block. Also, it is very difficult to obtain interpretable results as to how a city, composed of tens of isolated buildings of various heights, base areas and separations, responds to an earthquake, since the parametric studies this involves are necessarily numerous and extremely costly (thus rare) if this is done in a rigorous fashion.

Moreover, on account of the comments in the preceding section, it is not obvious whether  transfer functions at selected points (midpoints at ground floor or roof level of a building, midpoints between buildings, points on the ground far from the city center,....) and the associated response signals (notably the peak motion and duration \cite{tb75}, \cite{fvf90}, as well as other indicators \cite{aa15}, \cite{gbc02}, \cite{gtw05}, \cite{wg06a,wg06b}, \cite{h06}, \cite{sn15}  of the intensity of response at these points, furnish a sufficiently-complete picture of how and why an earthquake can provoke devastating effects in the structures of a city.

For these reasons, from here on, I choose to study, using the energy concepts illustrated in sect. \ref{khamen},  a city composed of a periodic assembly of homogenized, identical blocks (a block, contrary to a building is usually much larger in width than in height, and the successive blocks are assumed to be separated by streets). The issue of the homogenized nature of the blocks was discussed in sect. \ref{unicity}. The periodic nature of the arrangement of blocks was also justified by the empirical evidence that such is the urban pattern in areas of many modern cities. My study is 2D rather than 3D, for the obvious reasons of computational simplicity, but it would be very useful to find out if  the conclusions reached herein, apply, in their most important aspects, to 3D (and even non-periodic \cite{tb11a,tb11b}) urban patterns as well (the relative success of the ICBA formulation in describing energy communication to 2D city structures could make this a useful tool for studying the seismic response of 3D cities) .

The SFM and ICBA formulations can be applied indiscriminately to a city of buildings or to a city of blocks.  My cities of blocks are therefore very similar those of \cite{ks06} except as concerns the parameters $d$ (fixed), $w$ (fixed) and $h_{2}$ (variable):
\\\\
$c^{[0]''}=-5\approx 0~ms^{-1}$\\
$\mu^{[0]}=2\times 10^{9}~Pa$\\
$c^{[1]''}=4~ms^{-1}$\\
$\mu^{[1]}=7.2\times 10^{7}~ Pa$\\
$c^{[2]'}=240~ms^{-1}$\\
$c^{[2]''}=-12~ms^{-1}$\\
$\mu^{[2]}=14.4\times 10^{6}~ Pa$\\
$h_{[1]}=25~m$\\
$d=237.76~m$\\
$w=217.76~m$ (which means that the street width is $20~m$).
\clearpage
\newpage
\subsection{Variations of block height for typical city block}
Older cities, in so-called "less-developed countries", usually have less-tall buildings than those in newer cities of so-called "developed countries". A characteristic feature of the latter category of cities is that the height of many buildings therein is steadily increasing with time (due to increasing population concentration in cities \cite{b99}) so that the average height of a city block is also increasing . It is therefore of considerable interest to see what effect this produces on the way a city will react to an earthquake over a span of time \cite{kj06}, \cite{u10}.

\begin{figure}[ht]
  \begin{center}
    \subfloat[]{
      \includegraphics[width=0.5\textwidth]{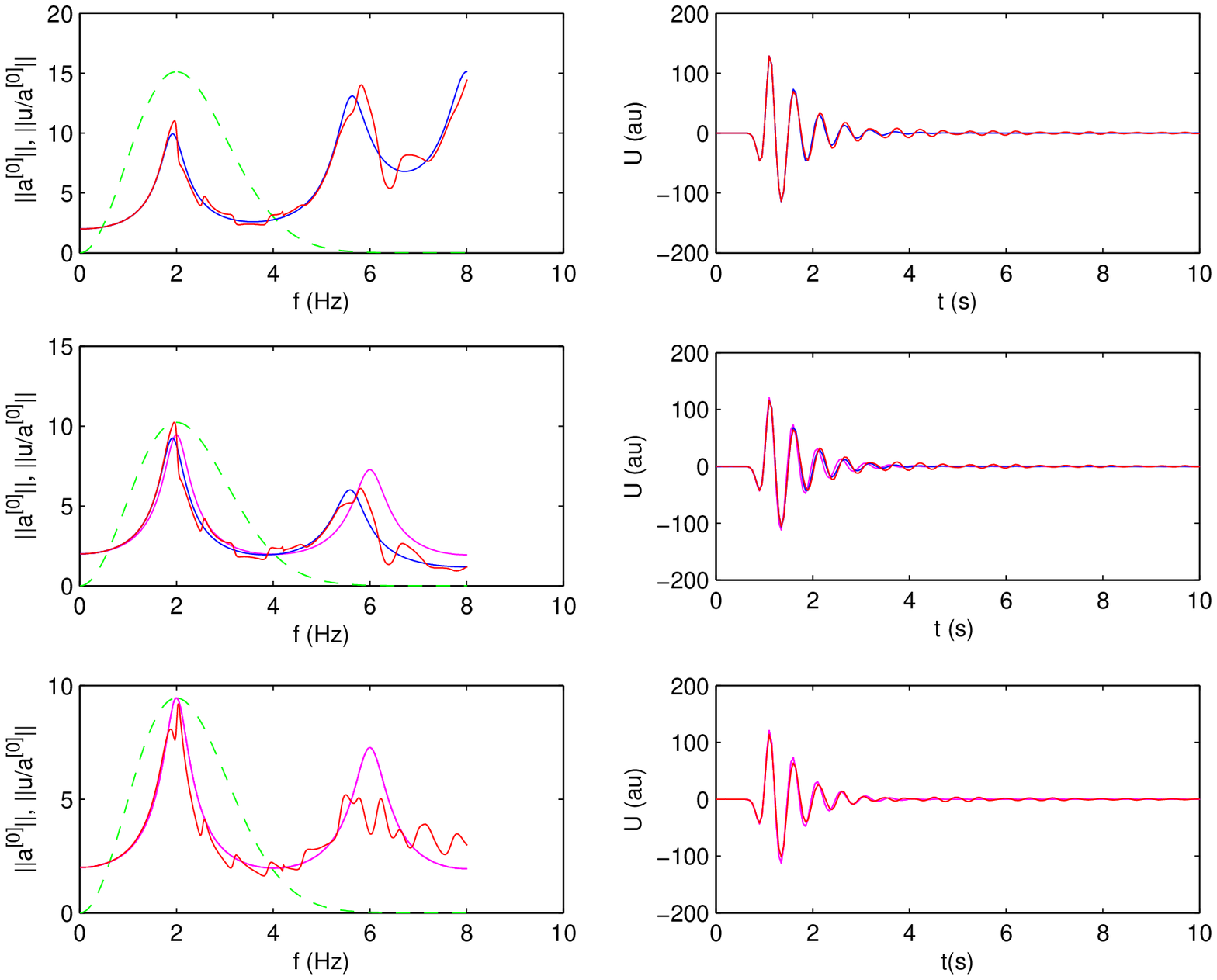}
      \label{a}
                         }
    \subfloat[]{
      \includegraphics[width=0.5\textwidth]{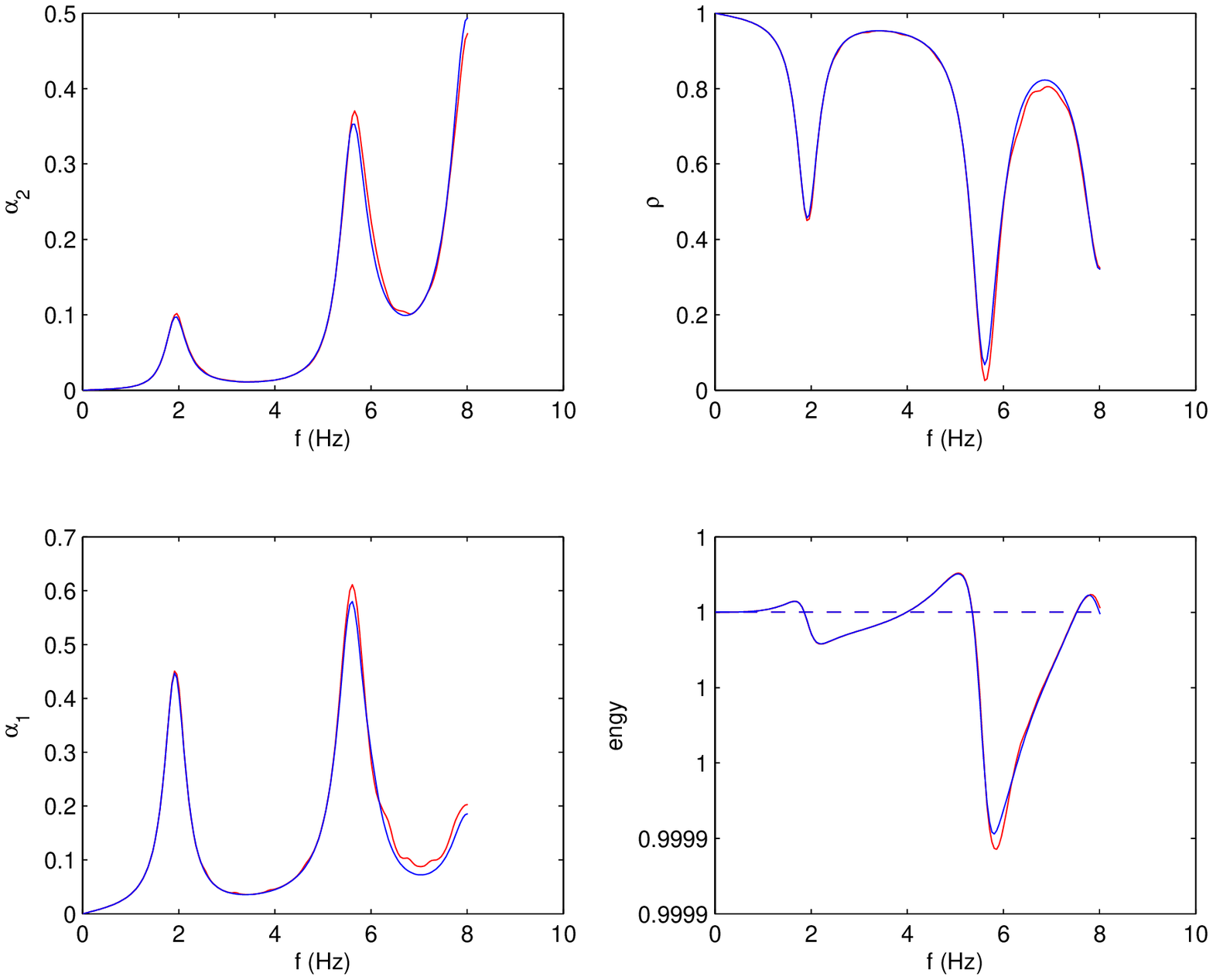}
      \label{b}
                         }
    \caption{
(a): Left-hand panels: modulus of transfer functions of displacement at the roof (top panels), base (middle panels), and on the ground in-betwwen buildings (bottom panels). Right-hand panels: corresponding time domain displacement responses.  $d=237.76~m$, $h_{1}=25~m$, $w=217.76~m$. $h_{2}=7.5~m$. $\nu=2~Hz,~\tau=1~s,~\theta^{i}=0^{\circ}$. \\
(b): The upper left-hand panel pertains to  the spectrum of normalized energy communicated to the generic building ($\alpha_{2}(\omega)$). The lower left-hand panel depicts the spectrum of normalized energy communicated to the soft layer ($\alpha_{1}(\omega))$. The upper right-hand panel pertains to the spectrum of energy communicated to the hard half space (i.e., radiation damping). The lower right-hand panel depicts the spectra of normalized input energy spectra (dashed curves) and normalized output energy(full curves). $d=237.76~m$, $h_{1}=25~m$, $w=217.76~m$. $h_{2}=7.5~m$.  $\theta^{i}=0^{\circ}$.
}
    \label{fks10}
  \end{center}
\end{figure}

\begin{figure}[ptb]
  \begin{center}
    \subfloat[]{
      \includegraphics[width=0.5\textwidth]{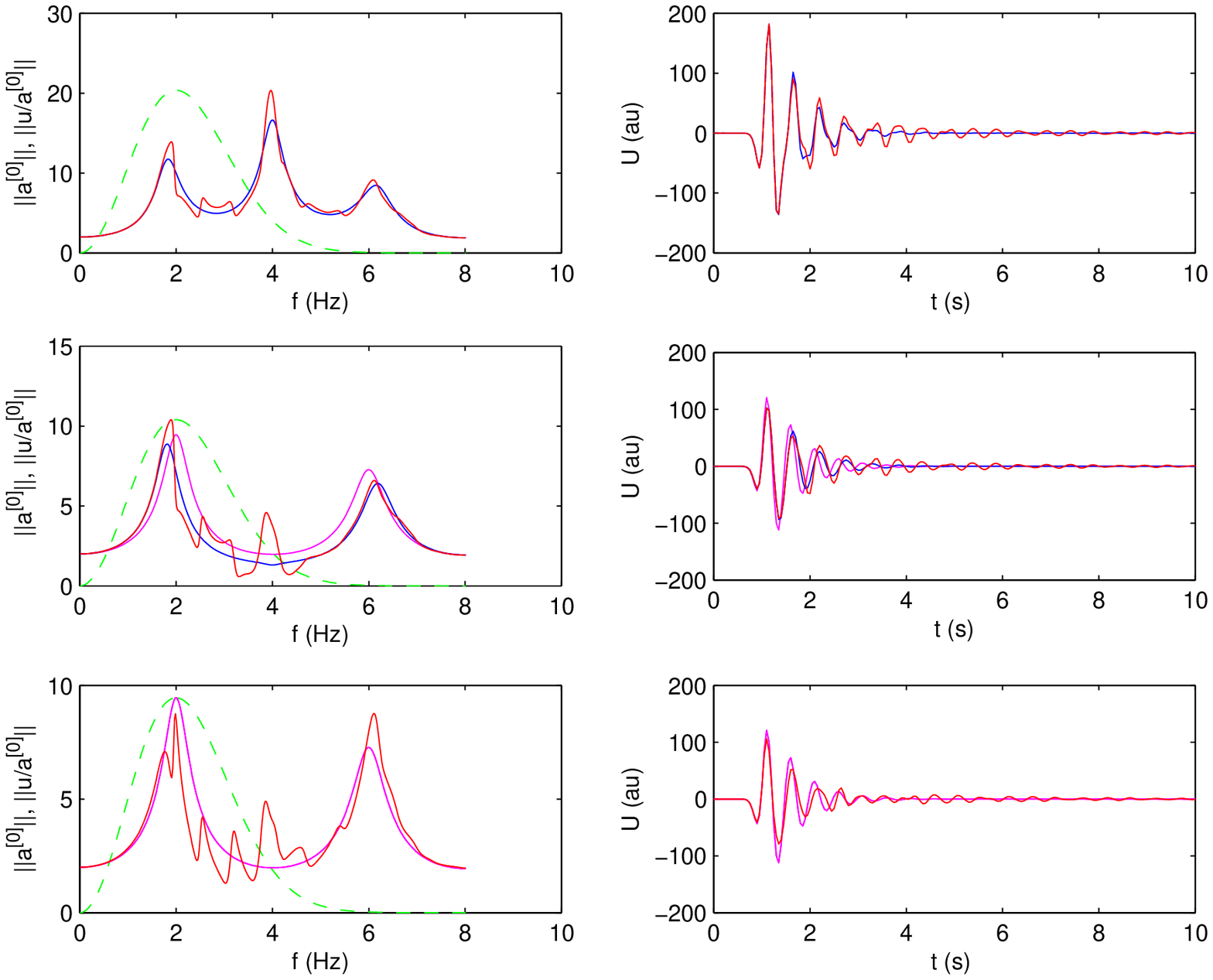}
      \label{a}
                         }
    \subfloat[]{
      \includegraphics[width=0.5\textwidth]{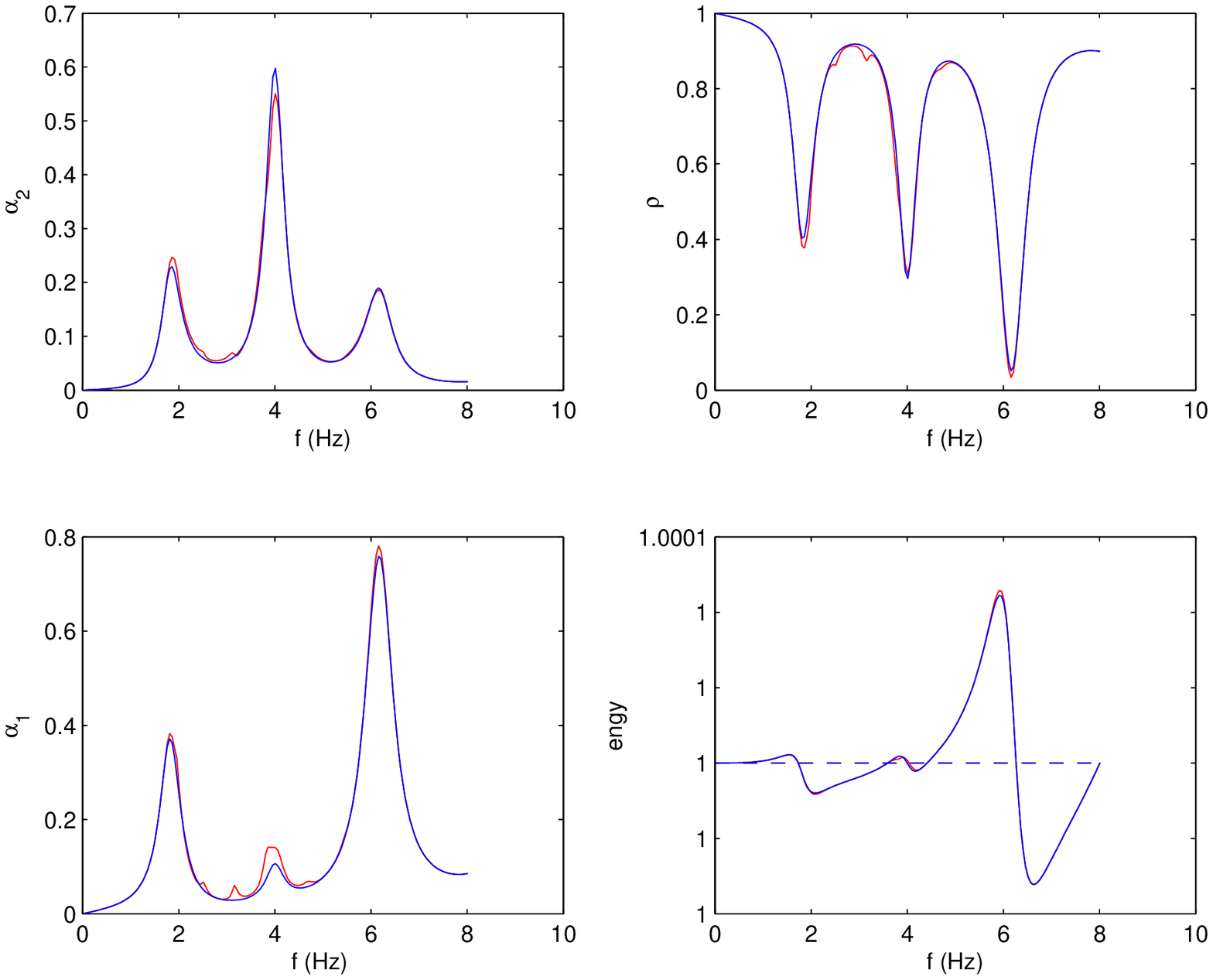}
      \label{b}
                         }
    \caption{
(a): Left-hand panels: modulus of transfer functions of displacement at the roof (top panels), base (middle panels), and on the ground in-betwwen buildings (bottom panels). Right-hand panels: corresponding time domain displacement responses.  $d=237.76~m$, $h_{1}=25~m$, $w=217.76~m$. $h_{2}=15~m$. $\nu=2~Hz,~\tau=1~s,~\theta^{i}=0^{\circ}$. \\
(b): The upper left-hand panel pertains to  the spectrum of normalized energy communicated to the generic building ($\alpha_{2}(\omega)$). The lower left-hand panel depicts the spectrum of normalized energy communicated to the soft layer ($\alpha_{1}(\omega))$. The upper right-hand panel pertains to the spectrum of energy communicated to the hard half space (i.e., radiation damping). The lower right-hand panel depicts the spectra of normalized input energy spectra (dashed curves) and normalized output energy(full curves).  $d=237.76~m$, $h_{1}=25~m$, $w=217.76~m$. $h_{2}=15~m$.  $\theta^{i}=0^{\circ}$.
}
    \label{fks20}
  \end{center}
\end{figure}

\begin{figure}[ptb]
  \begin{center}
    \subfloat[]{
      \includegraphics[width=0.5\textwidth]{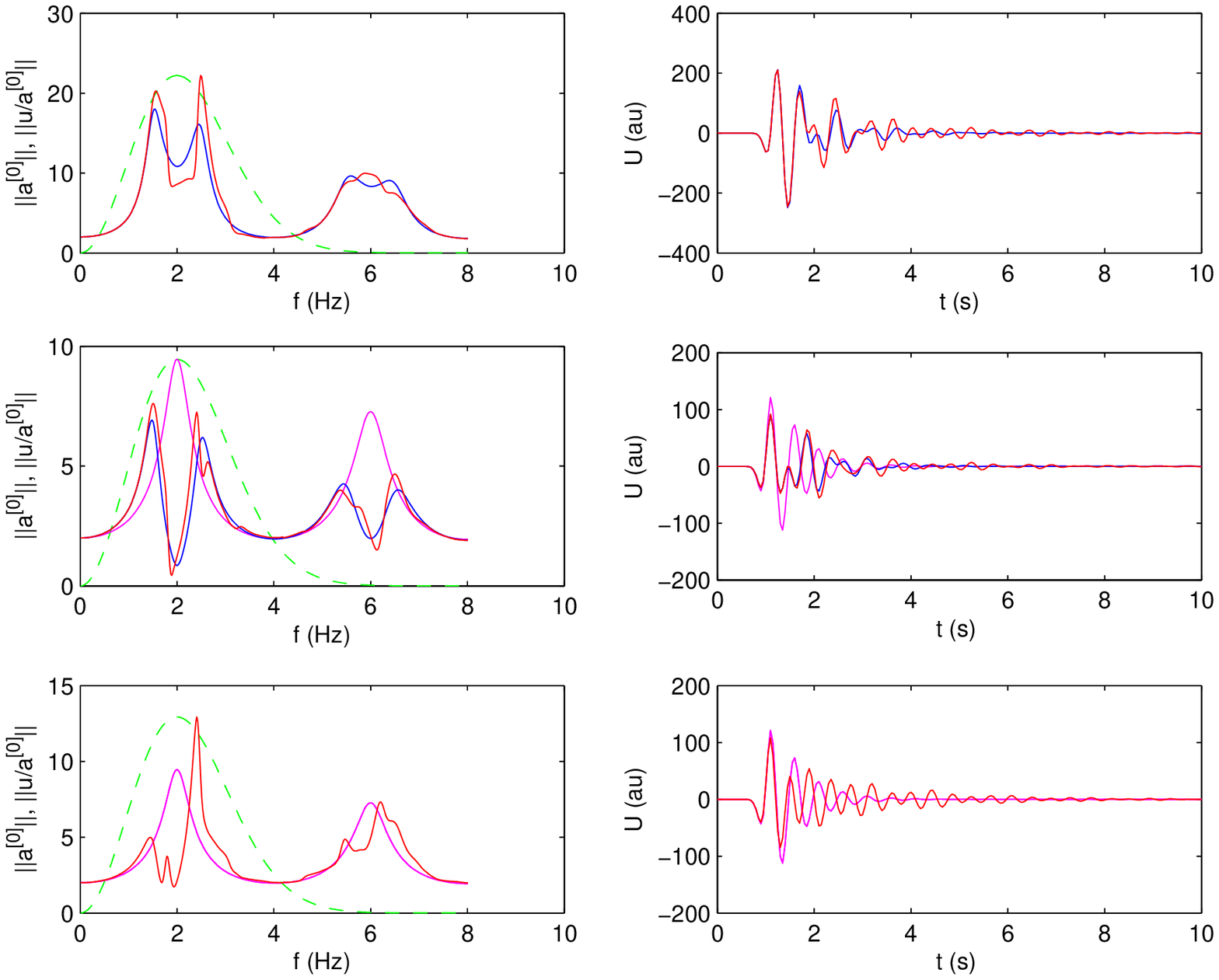}
      \label{a}
                         }
    \subfloat[]{
      \includegraphics[width=0.5\textwidth]{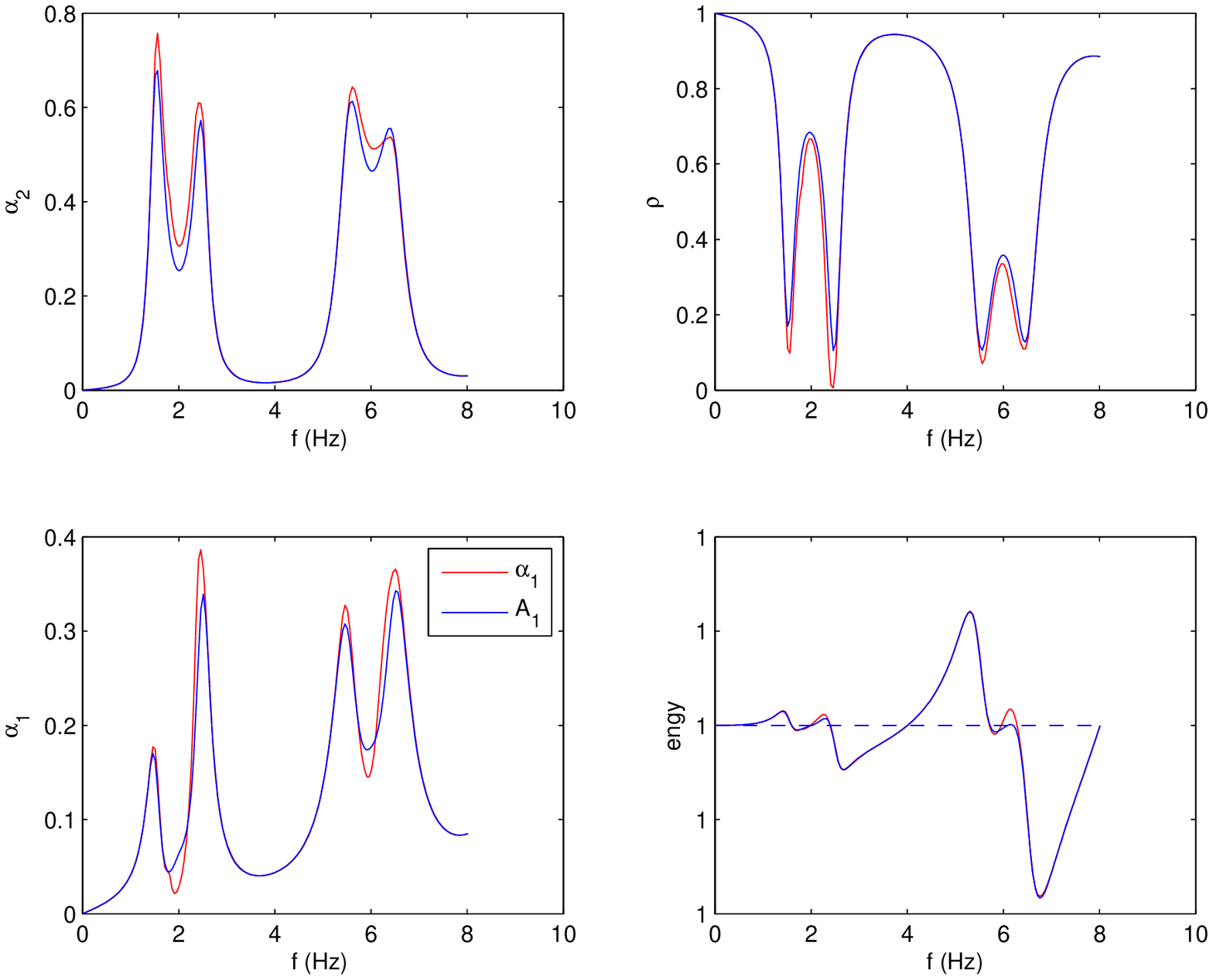}
      \label{b}
                         }
    \caption{
(a): Left-hand panels: modulus of transfer functions of displacement at the roof (top panels), base (middle panels), and on the ground in-betwwen buildings (bottom panels). Right-hand panels: corresponding time domain displacement responses.  $d=237.76~m$, $h_{1}=25~m$, $w=217.76~m$. $h_{2}=30~m$. $\nu=2~Hz,~\tau=1~s,~\theta^{i}=0^{\circ}$. \\
(b): The upper left-hand panel pertains to  the spectrum of normalized energy communicated to the generic building ($\alpha_{2}(\omega)$). The lower left-hand panel depicts the spectrum of normalized energy communicated to the soft layer ($\alpha_{1}(\omega))$. The upper right-hand panel pertains to the spectrum of energy communicated to the hard half space (i.e., radiation damping). The lower right-hand panel depicts the spectra of normalized input energy spectra (dashed curves) and normalized output energy(full curves).  $d=237.76~m$, $h_{1}=25~m$, $w=217.76~m$. $h_{2}=30~m$.  $\theta^{i}=0^{\circ}$.
}
    \label{fks30}
  \end{center}
\end{figure}

\begin{figure}[ptb]
  \begin{center}
    \subfloat[]{
      \includegraphics[width=0.5\textwidth]{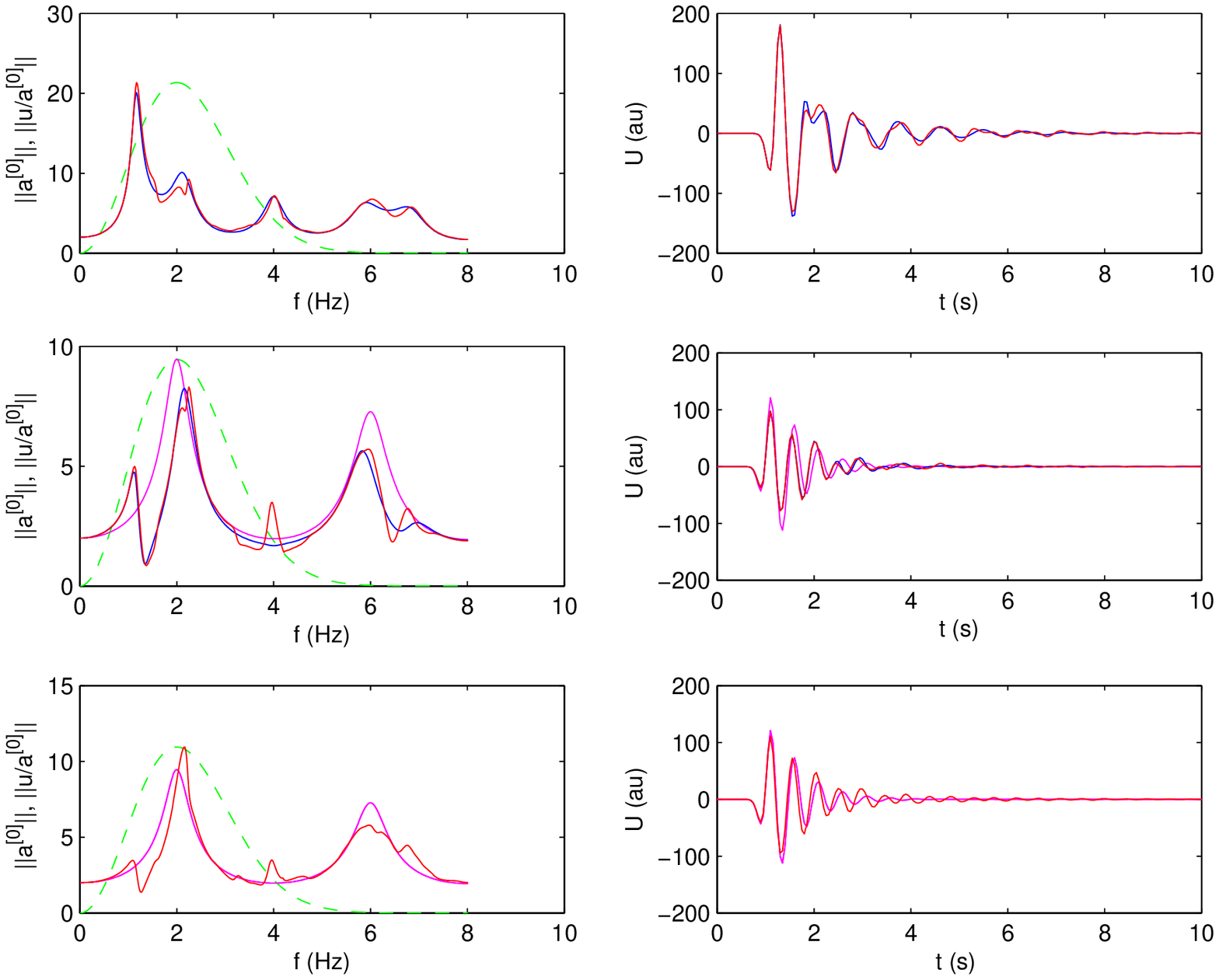}
      \label{a}
                         }
    \subfloat[]{
      \includegraphics[width=0.5\textwidth]{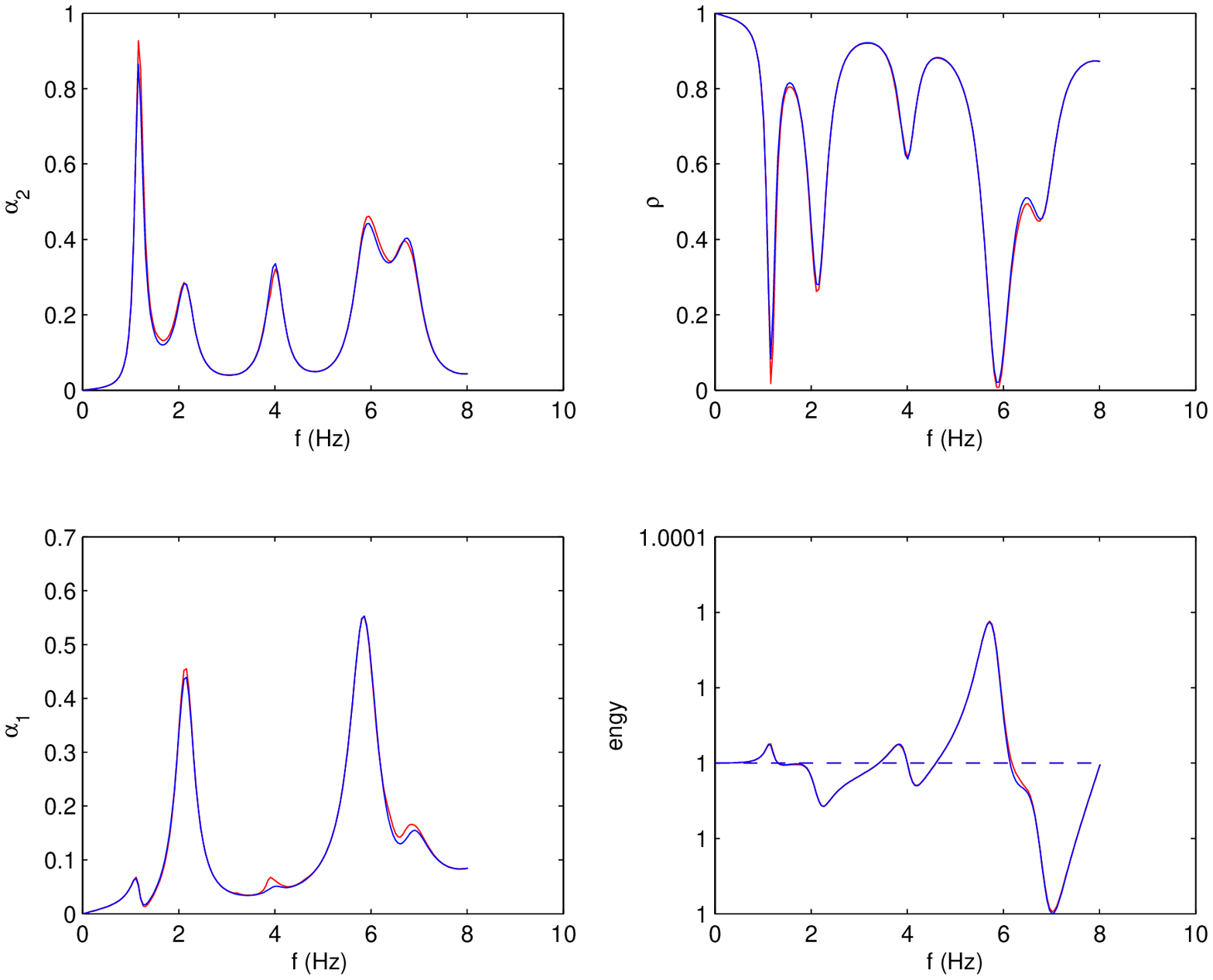}
      \label{b}
                         }
    \caption{
(a): Left-hand panels: modulus of transfer functions of displacement at the roof (top panels), base (middle panels), and on the ground in-betwwen buildings (bottom panels). Right-hand panels: corresponding time domain displacement responses.  $d=237.76~m$, $h_{1}=25~m$, $w=217.76~m$. $h_{2}=45~m$. $\nu=2~Hz,~\tau=1~s,~\theta^{i}=0^{\circ}$. \\
(b): The upper left-hand panel pertains to  the spectrum of normalized energy communicated to the generic building ($\alpha_{2}(\omega)$). The lower left-hand panel depicts the spectrum of normalized energy communicated to the soft layer ($\alpha_{1}(\omega))$. The upper right-hand panel pertains to the spectrum of energy communicated to the hard half space (i.e., radiation damping). The lower right-hand panel depicts the spectra of normalized input energy spectra (dashed curves) and normalized output energy(full curves).  $d=237.76~m$, $h_{1}=25~m$, $w=217.76~m$. $h_{2}=45~m$.  $\theta^{i}=0^{\circ}$.
}
    \label{fks40}
  \end{center}
\end{figure}

Figs. \ref{fks10}-\ref{fks40} are each composed of two subfigures denoted by (a) and (b). Each left-hand subfigure is relative to the transfer functions and temporal responses, whereas each right-hand subfigure describes $\alpha_{1}$ (relative energy communicated to the layer), $\alpha_{2}$ (relative energy communicated to the city), $\rho$ (relative energy carried away in the underlying half space in the form of radiation damping), $E_{out}=\alpha_{1}+\alpha_{2}+\rho$ (relative output energy), knowing that $E_{out}=E_{in}$ ($E_{in}=1$=relative input energy) for energy to be conserved. The color schemes are the same as previously.
\clearpage
\newpage
Fig. \ref{fks10} corresponds to a present-day city ($h_{2}=7.5~m$) such as Port-au-Prince (Haiti) or Kathmandu (Nepal) . The energy communicated to this city is  relatively small in the neighborhood of the (first) soil frequency ($2~Hz$, as previously), but, of course, this small energy can destroy entire blocks of poorly-built buildings (which is what happened in 2010 in Port-au-Prince  and in 2015 in Kathmandu)  if the spectrum of the seismic pulse peaks near the soil frequency. Interestingly, if the spectrum peaks near the second soil frequency, the energy communicated to the blocks is over four times what it was in the former situation.

Fig. \ref{fks20} corresponds to the early epoch of  some downtown areas of a city such as San Francisco ($h_{2}=15~m$). The energy communicated to this city is  twice as large as previously in the neighborhood of $\approx 2~Hz$, which  would have been sufficient to damage or destroy a  great deal of the structures in areas of San Francisco during the 1906  earthquake had the seismic pulse peaked near $2~Hz$. If, on the contrary, it had peaked near the $\approx 4~Hz$  then  large-scale damage and destruction would have been even more likely.

Fig. \ref{fks30} corresponds to the  situation of  many areas of a city such as Mexico City (in 1985),  Kobe (Japan) or Sendai (Japan) ($h_{2}=30~m$) at the time at which they were struck by large-magnitude earthquakes. Now, two split, rather wideband, features of strong communication of energy to the city blocks are observed to exist around $2~Hz$ and $6~Hz$. It is known \cite{rg12} that mainshock-aftershock sequences are often characterized by relatively-low frequency and relatively-high frequency spectra respectively, which means that if the peaks of the mainshock-aftershock sequences in Mexico City (1985) (and Concepcion (2010)) were at $2~Hz$ and $6~Hz$ respectively then the $30~m$ high blocks   would have been strongly affected by the mainshock but also, and as strongly, by the aftershock.

Fig. \ref{fks40} corresponds to the  situation of  certain areas of a city such as Mexico City (at present) and Tainan (at the time of the  2016 earthquake) ($h_{2}=45~m$). Now the first split peaks are narrower and more intense, which for the lowest frequency peak corresponds to very large communication of energy into blocks containing predominantly-tall buildings. The other peaks around $2~Hz$, $4~Hz$ and $6~Hz$ are  weaker, but could be responsible for supplementary damage due to aftershocks \cite{k12}.

An important sidelight of figs. \ref{fks20}-\ref{fks40} is that the ICBA is seen to agree quite well with the SFM formulation in predicting the distribution of energy in a city of blocks.
\subsection{Variations of the angle of incidence for typical city block}
\begin{figure}[h]
\begin{center}
      \includegraphics[width=0.85\textwidth]{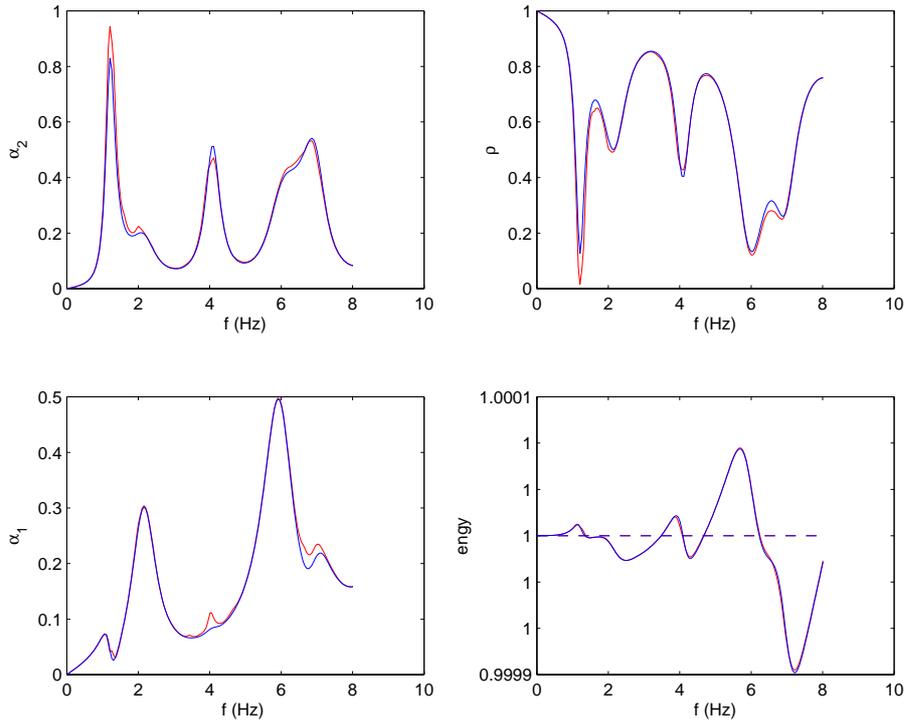}
    \caption{The upper left-hand panel pertains to  the spectrum of normalized energy communicated to the generic building ($\alpha_{2}(\omega)$). The lower left-hand panel depicts the spectrum of normalized energy communicated to the soft layer ($\alpha_{1}(\omega))$. The upper right-hand panel pertains to the spectrum of energy communicated to the hard half space (i.e., radiation damping). The lower right-hand panel depicts the spectra of normalized input energy  (dashed curves) and normalized output energy (full curves). $d=237.76~m$, $h_{1}=25~m$, $w=217.76~m$. $h_{2}=45~m$.  $\theta^{i}=60^{\circ}$.}
    \label{fks60}
  \end{center}
\end{figure}

\begin{figure}[h]
\begin{center}
      \includegraphics[width=0.85\textwidth]{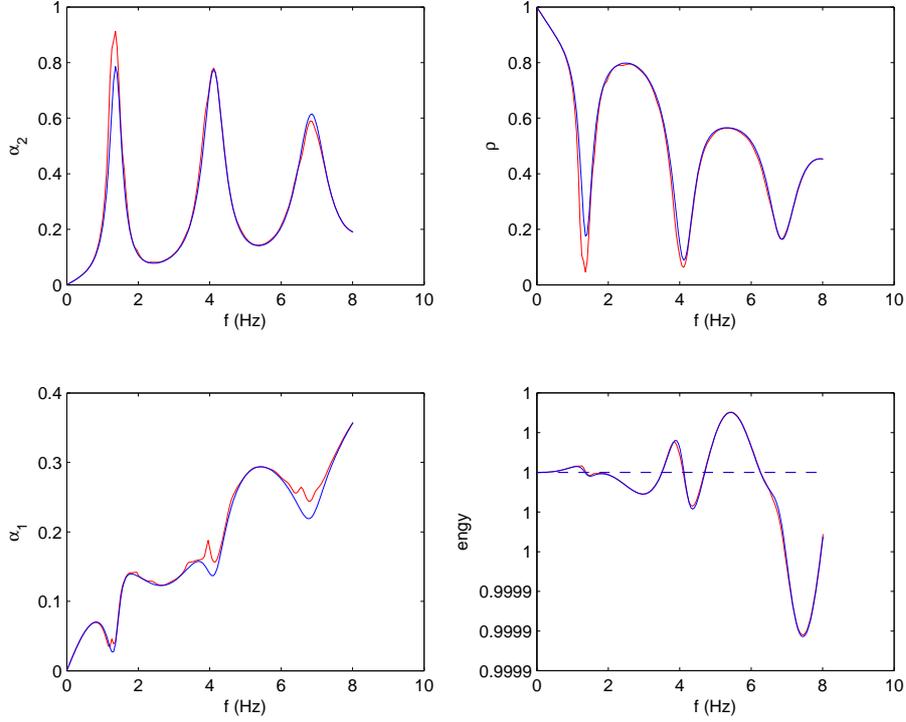}
    \caption{The upper left-hand panel pertains to  the spectrum of normalized energy communicated to the generic building ($\alpha_{2}(\omega)$). The lower left-hand panel depicts the spectrum of normalized energy communicated to the soft layer ($\alpha_{1}(\omega))$. The upper right-hand panel pertains to the spectrum of energy communicated to the hard half space (i.e., radiation damping). The lower right-hand panel depicts the spectra of normalized input energy  (dashed curves) and normalized output energy (full curves). $d=237.76~m$, $h_{1}=25~m$, $w=217.76~m$. $h_{2}=45~m$.  $\theta^{i}=80^{\circ}$.}
    \label{fks70}
  \end{center}
\end{figure}

In most studies of earthquakes in cities, it is assumed that the seismic disturbance takes the form of a vertically-incident (i.e., $\theta^{i}=0^{\circ}$,  the seismic source is laterally very near to, and vertically very deep under, the city) plane bulk wave. However, it is a well-documented fact \cite{j89}, \cite{rg12}, \cite{kj93}, \cite{c14}, \cite{cscg03}, \cite{cgass89}, \cite{cs00}, \cite{me06}) that many of the deadliest earthquakes in cities were due to  laterally-far, and vertically-relatively-shallow, sources. To see how this affects the seismic response in the city, I simply chose incident angles $\theta^{i}$ different from $0^{\circ}$ \cite{t72}, \cite{lw82}, \cite{dg07} (the authors of \cite{ll12} even replace the plane wave by the wave radiated by a finite-fault source, but their site has no soft layer). The results (not unlike those of \cite{lw82}) are presented in figs. \ref{fks60} ($\theta^{i}=60^{\circ}$)  and \ref{fks70} ($\theta^{i}=80^{\circ}$) which should be compared to the previous fig. \ref{fks40} ($\theta^{i}=0^{\circ}$).
\clearpage
\newpage
It is seen from these figures that the incident angle has relatively-little effect: (a) on the quantity of energy communicated to the blocks, and (b) on the location of the spectral bands in which this energy absorption occurs, so that theory is consistent with empirical evidence concerning  the fact that  earthquakes due to  laterally very-distant sources are also able to produce considerable damage in cities.

An important sidelight of figs. \ref{fks60}-\ref{fks70} is again that the ICBA is seen to agree quite well with the SFM formulation in predicting the distribution of energy in a city of blocks, even for oblique incidence of the seismic wave.
\clearpage
\newpage
\section{Conclusions and perspectives}
One would think that the importance of the subject of seismic motion in cities  be proportional to the rate of increase (which is explosive, particularly in China) of the number and size  of earthquake-prone cities \cite{b99}. Curiously, the amount of research devoted to this subject is not only modest, but not increasing in proportion to its importance. The purpose of the present contribution was to hopefully rekindle interest in this scientifically-fascinating, and awesome, by its social and economic implications, subject.

Research on seismic motion in cities is  in need of new concepts and methods of analysis. It was proposed herein: 1) that a city be treated as another geological layer on top of the soil layer, 2) and/or that the city be viewed as  a periodic distribution of low aspect-ratio blocks,  and 3) that the central issue be the energy communicated to the buildings or blocks of the city. The proposed  methods of analysis were: 1) focusing attention on the conservation and distribution of energy within the different components of the city site rather than on ground motion, and 2) the ICBA, as a means of simplifying, or suggesting new angles of attack to, a global analysis of the problem.

The approximate formulation constituted by the ICBA was confronted, both quantitatively with the rigorous SFM {\cite{g05,gw08} and boundary element formulations  \cite{ks06} (notably as concerns ground motion), and qualitatively with the approximate equivalent surface impedance formulation \cite{wg06a,wg06b,sb16} (notably as concerns the phenomenon of transfer function peak splitting provoked by the presence of a city of sufficient height). It turned out that the ICBA yields quantitatively-exact descriptions of both ground and building motion when the city is viewed either as a uniform layer or as an assembly of relatively-low aspect ratio blocks, and also gives rise to qualitatively-interesting descriptions even for cities of high aspect ratio buildings or blocks.

The conserved-energy analysis enabled a quantitative description of how the energy is distributed, as a function of the frequency of the seismic load,  to the various regions (half space, soil layer, and city layer, blocks, or buildings) of the city site. It was thus shown: (a) that in certain frequency intervals, the energy communicated to a city block as a whole, and even to a single building thereof, can be large even when the ground motion is smaller than what it would be in the absence of the block, and (b)  that a substantial amount of energy can be transferred to the buildings or blocks even at the second building or block frequency, which fact might give insight as to how mainshock-aftershock sequences are able to aggravate the damage inflicted on cities.

It was also found that theory is consistent with empirical evidence concerning  the fact that  earthquakes  are  able to communicate large amounts of energy and therefore cause  considerable damage to buildings even when the seismic source is   laterally very-distant from the city.

Finally, it was demonstrated, both theoretically and numerically that, although the so-called "beneficial effect" (manifested by reduced ground motion at the frequency of coincidence of the soil and building frequencies) of the   added mass of buildings on the ground really exists, it does not translate to "reduced motion" within the buildings or groups of buildings of a city, because, in fact, at or near the coincidence frequency the energy transferred to the buildings can attain of the order of one half of the input seismic energy.

There remain many unanswered questions, the foremost of which concerns long-duration earthquakes in urban centers such as Mexico City. In the examples of the present paper, the durations of the shaking in the buildings or (connected or disconnected) blocks did not exceed $10~s$ whereas what has been recorded in Mexico City \cite{cgass89,cscg03} and Sendai \cite{c08,c14,k12} are strong vibrations that last {\it a few minutes}. A clue to why this can occur has been provided in \cite{wg06a,wg06b} concerning the factors that favor the coupling of seismic energy from its source to Love waves (which, like the Rayleigh waves, are  manifestations of authentic resonant modes, in contrast to the interference waves of the type excited in horizontally-layered systems by plane waves that are only pseudo-resonant modes \cite{g05,gw08}), and therefore to ground (not loaded by added mass) responses of increased duration;  this indication should give rise to a more global, although preliminary, study, employing the uniform city model in conjunction with the ICBA or boundary element methods of analysis.

Also, the ICBA should be compared to the boundary element method, and/or perhaps the equivalent surface impedance method, and then be applied to either plane wave or source wave loading in a full 3D setting, while employing the energy distribution concepts developed herein. This could provide an answer to  the question of whether radiation damping might be increased in 3D configurations and thus potentially lessen the damage inflicted to buildings.

Finally, a question of great importance concerns the homogenizations implicit in the uniform layer, uniform block, uniform building, equivalent surface impedance models, which are necessarily-invoked to simulate seismic motion in cities with a reasonable amount of computations: how to best account for the response (average or otherwise) of the complicated structures that are the layer, block or building with the least amount of parameters and for  many types of solicitations (in terms of polarization, frequency, and spatial characteristics of the source)?

\end{document}